**Moving from two- to multi-way interactions among binary risk factors on the additive scale.**

Michail Katsoulis[1,2], Christina Bamia[†]

1. Institute of Health Informatics, University College London, London, UK
2. Hellenic Health Foundation, Athens, Greece

**Corresponding Author**

Michail Katsoulis

Institute of Health Informatics, University College London, London, UK

Tel.: +44 (0) 203 549 5022

Email: m.katsoulis@ucl.ac.uk

[†]deceased

## ABSTRACT

Many studies have focused on investigating deviations from additive interaction of two dichotomous risk factors on a binary outcome. There is, however, a gap in the literature with respect to interactions on the additive scale of >2 risk factors. In this paper, we present an approach for examining deviations from additive interaction among three on more binary exposures. The relative excess risk due to interaction (RERI) is used as measure of additive interaction. First, we concentrate on three risk factors - we propose to decompose the total RERI to: the RERI owned to the joint presence of all 3 risk factors and the RERI of any two risk factors, given that the third is absent. We then extend this approach, to >3 binary risk factors. For illustration, we use a sample from data from the Greek EPIC cohort and we investigate the association with overall mortality of Mediterranean diet, body mass index (BMI), and, smoking. Our formulae enable better interpretability of any evidence for deviations from additivity owned to more than two risk factors and provide simple ways of communicating such results from a public health perspective by attributing any excess relative risk to specific combinations of these factors.

## INTRODUCTION

Four decades ago, Rothman stated that, as more than one risk factors are eventually established for the etiology of a specific health outcome, epidemiologists will need to pay more attention to the issue of interaction (synergy or antagonism) between these factors[1]. This is particularly relevant in the field of genetic epidemiology, as scientists focus on the study of thousands of genes and of their interactions with environmental factors[2-3].

Measuring interaction on the additive scale is more important from a public health perspective[4-7], because, in this context, two risk factors are independent, when the number of disease cases is not dependent on the extent to which these factors act together[5]. If, for example, the number (or rate) of hospitalizations for a disease when individuals are exposed into two risk factors is greater than the sum of hospitalizations for this disease of the people exposed only to one of these factors, then the public health services would be challenged to carry extra weight due to this interaction, which is measured as a deviation from additivity of the effects of these factors.

Nevertheless, the usual practice has been to refer to statistical interaction when studying interaction between risk factors[7]. Under this concept, interaction is measured on either additive or multiplicative scale, depending solely on the form of the underlying model used, rather than on a-priori consideration for the expected type of associations between these exposures and the outcome.

The deviation from additivity of the effects between two variables has been proposed by Rothman[8] and further explored by others[2-3,9-22]. Surprisingly, the study of additive interaction of >2 factors has not been studied adequately, even if it would enable better understanding of the joint action of many factors for the development of a specific disease. This may have occurred because conceptualizing the features of multi-way interaction is challenging (e.g. a modification of an interaction between two variables by a third variable is not easy to understand and sometimes the rationale for assessing this effect is not there). Apart from the studies related to multi-way interaction in the sufficient-cause framework[23-25], to the best of our knowledge, there are only three relevant publications that have focused only on practical illustration of deviation from additivity of the effects of three risk factors[26-28].

In this paper, we aim to fill this gap in the literature and provide useful tools for researchers who wish to focus on the joint action of >2 binary factors. We highlight the questions of interest in the study of joint action of 3 factors, we give answers by introducing useful indexes for additive interaction, accompanied by the corresponding recommendation for the conduction these of analyses, and we then extend the methods to >3 factors. We illustrate our theoretical arguments using data from the Greek-EPIC study and we provide an easy-to-use code in Stata for the implementation of these methods.

## METHODS

### *Definitions*

Consider n dichotomous $X_i$ , i=1,..,n variables as risk factors for a disease D with $X_i$ = (0,1). Let D+ and D- denote the presence/absence of D, and $X_i$+, $X_i$-, the presence ($X_i$ =1) or absence ($X_i$ =0) of $X_i$. The relative risk of D+ for any combination of the presence or absence of $X_1$, $X_2$, ... $X_n$, as compared to their absence is denoted by $RR_{X_1\#X_2\#\ldots X_n\#}$ , where # = +/- and the corresponding excess relative risk by $ERR_{X_1\#X_2\#\ldots X_n\#}$ with

$ERR_{X_1\#X_2\#\ldots X_n\#} = RR_{X_1\#X_2\#\ldots X_n\#} - RR_{X_1-X_2-\cdots X_n-}$ , i.e.

$$\mathrm{ERR}_{X_1\#X_2\#\ldots X_n\#} = \mathrm{RR}_{X_1\#X_2\#\ldots X_n\#} - 1$$

For the investigation of any deviation from additivity of the effects of two risk factors, we focus on the contrast between

$$\mathrm{RR}_{X_1+X_2+} - \mathrm{RR}_{X_1-X_2-} \text{ vs } \left(\mathrm{RR}_{X_1+X_2-} - \mathrm{RR}_{X_1-X_2-}\right) - \left(\mathrm{RR}_{X_1-X_2+} - \mathrm{RR}_{X_1-X_2-}\right)$$

i.e. the excess risk from the situation when two risk factors act jointly versus the extra risk of the occasions that each of them acts separately

So a measure for additive interaction would be the relative excess risk due to interaction $RERI_2(X_1,X_2)$

$$\begin{aligned}\mathrm{RERI}(X_1, X_2) &= \mathrm{ERR}_{X_1+X_2+} - \mathrm{ERR}_{X_1+X_2-} - \mathrm{ERR}_{X_1-X_2+} \\ &= \left(\mathrm{RR}_{X_1+X_2+} - \mathrm{RR}_{X_1-X_2-}\right) - \left(\mathrm{RR}_{X_1+X_2-} - \mathrm{RR}_{X_1-X_2-}\right) - \left(\mathrm{RR}_{X_1-X_2+} - \mathrm{RR}_{X_1-X_2-}\right) \\ &= \mathrm{RR}_{X_1+X_2+} - \mathrm{RR}_{X_1+X_2-} - \mathrm{RR}_{X_1-X_2+} + 1\end{aligned}$$

which indicates whether the effect of 2 risk factors that act jointly is greater ($RERI_2>0$), equal ($RERI_2=0$) or lower ($RERI_2<0$) than the sum of their individual effect (super-additive, additive or sub-additive effects respectively).

It is crucial to highlight that the factors cannot be protective, because the calculation of additive interaction will be wrong, as a relative risk is between 0 and 1 for a protective factor, while it can be from 1 to infinity for a risk factor. Imagine 2 drugs with additive effects ($RERI_2=0$) on CVD, each of those reducing the CVD risk by 75% (i.e. $RR_{10}$,$RR_{01}=0.25$). We cannot use the $RERI_2$ index, because we would calculate that $RR_{11}$ is negative ($RR_{11}=–0.\ 5$)! Instead, we should recode these factors into risk (i.e. the effect of not taking the drugs) and apply the calculations (see [9] and Appendix, Section B and D).

Additionally, if one wants to focus on the multiplicative interaction of 2 risk factors, then the contrast of interest would be $\mathrm{RR}_{\mathrm{X}_1+\mathrm{X}_2+}$ vs $\mathrm{RR}_{\mathrm{X}_1+\mathrm{X}_2-} * \mathrm{RR}_{\mathrm{X}_1-\mathrm{X}_2+}$

i.e. the comparison of the relative risk when two risk factors act jointly versus the multiplication of the risk of the occasions that each of them acts separately

The index of multiplicative interaction would be

$$I_2 = \frac{\mathrm{RR}_{\mathrm{X}_1+\mathrm{X}_2+}}{\mathrm{RR}_{\mathrm{X}_1+\mathrm{X}_2-} * \mathrm{RR}_{\mathrm{X}_1-\mathrm{X}_2+}}$$

For more details, see ref [16].

In Appendix (Section E), we show i) that multiplicative or super-multiplicative effects imply super-additive effects and ii) that additive or sub-additive effects imply sub-multiplicative effects for 2-way interactions.

*From the two to the three-way interaction on the additive scale*

Imagine now that the question of interest is whether 3 risk factors “interact” on the additive scale. How should we face that problem?

The first answer we should give would be an extension of the previous methods for the construction of $\mathrm{RERI}_2$. Now, we should take into account the extra risk due to the joint presence of the three risk factors and compare it with the sum of the excess risks caused by each risk factor separately, i.e.

$$\left(\mathrm{RR}_{\mathrm{X}_1+\mathrm{X}_2+\mathrm{X}_3+} - \mathrm{RR}_{\mathrm{X}_1-\mathrm{X}_2-\mathrm{X}_3-}\right) \text{ vs } \left(\mathrm{RR}_{\mathrm{X}_1+\mathrm{X}_2-\mathrm{X}_3-} - \mathrm{RR}_{\mathrm{X}_1-\mathrm{X}_2-\mathrm{X}_3-}\right) + \left(\mathrm{RR}_{\mathrm{X}_1-\mathrm{X}_2+\mathrm{X}_3-} - \mathrm{RR}_{\mathrm{X}_1-\mathrm{X}_2-\mathrm{X}_3-}\right) + \left(\mathrm{RR}_{\mathrm{X}_1-\mathrm{X}_2-\mathrm{X}_3+} - \mathrm{RR}_{\mathrm{X}_1-\mathrm{X}_2-\mathrm{X}_3-}\right)$$

In other words, we should extend the RERI definition to three risk factors $X_1$, $X_2$ and $X_3$, and calculate the total relative excess risk due to interaction ($\mathrm{TotRERI}_3$) ,

$$\mathrm{TotRERI}_3(\mathrm{X}_1, \mathrm{X}_2, \mathrm{X}_3) = \left(\mathrm{RR}_{\mathrm{X}_1+\mathrm{X}_2+\mathrm{X}_3+} - \mathrm{RR}_{\mathrm{X}_1-\mathrm{X}_2-\mathrm{X}_3-}\right) - \left(\mathrm{RR}_{\mathrm{X}_1+\mathrm{X}_2-\mathrm{X}_3-} - \mathrm{RR}_{\mathrm{X}_1-\mathrm{X}_2-\mathrm{X}_3-}\right)$$

$$-\left(RR_{X_1-X_2+X_3-} - RR_{X_1-X_2-X_3-}\right) - \left(RR_{X_1-X_2-X_3+} - RR_{X_1-X_2-X_3-}\right)$$

$$= RR_{X_1+X_2+X_3+} - RR_{X_1+X_2-X_3-} - RR_{X_1-X_2+X_3-} - RR_{X_1-X_2-X_3+} + 2 \qquad (1)$$

The total relative excess risk due to interaction ($TotRERI_3$) is calculated by comparing the joint effect of three risk factors to the situation when each one acts separately. It allows us to understand whether these variables have super-additive, additive or sub-additive effects.

The next issue that we should wonder is about the index for 3-way additive interaction, beyond two-way interactions. The super/sub additivity of the effects of 3 risk factors (1) is due either to the 3-way interaction ($RERI_3$) of the 3 risk factors, or to the 2-way interaction of the 2 risk factors, when the 3rd is absent. To calculate $RERI_3$, one needs to subtract $RERI_2(X_1,X_2 \mid X_3=0)$, $RERI_2(X_1,X_3 \mid X_2=0)$ and $RERI_2(X_2,X_3 \mid X_1=0)$ from $TotRERI_3$, i.e.

$$RERI_3(X_1, X_2, X_3) = TotRERI_3(X_1, X_2, X_3)$$

$$-RERI_2(X_1, X_2|X_3 = 0) - RERI_2(X_1, X_3|X_2 = 0) - RERI_2(X_2, X_3|X_1 = 0)$$

The relative excess risk due to interaction is the measure of the three-way interaction. This index indicates whether there is positive/negative 3-way interaction on the additive scale, which is explicitly due to the joint presence of all three factors, in other words, measures the 3-way interaction, beyond the possible 2 way interactions. Additionally, , $TotRERI_3(X_1, X_2, X_3)$ expresses the sum of the 3-way interaction and all 2-way interactions and , that is

$$TotRERI_3(X_1, X_2, X_3) = RERI_3(X_1, X_2, X_3)$$

$$+RERI_2(X_1, X_2|X_3 = 0) + RERI_2(X_1, X_3|X_2 = 0) + RERI_2(X_2, X_3|X_1 = 0)$$

(2)

Of note, $TotRERI_3$ may be zero as the result of 2-way interactions that cancel out with 3-way interaction.

Moreover, to calculate the three-way interaction ($\mathrm{RERI}_3(\mathrm{X}_1, \mathrm{X}_2, \mathrm{X}_3)$),we have to combine (1) and (2) (see Appendix, section A) to estimate

$$\mathrm{RERI}_3(\mathrm{X}_1, \mathrm{X}_2, \mathrm{X}_3) = \mathrm{RR}_{\mathrm{X}_1+\mathrm{X}_2+\mathrm{X}_3+} \\ -\mathrm{RR}_{\mathrm{X}_1+\mathrm{X}_2+\mathrm{X}_3-} - \mathrm{RR}_{\mathrm{X}_1+\mathrm{X}_2-\mathrm{X}_3+} - \mathrm{RR}_{\mathrm{X}_1-\mathrm{X}_2+\mathrm{X}_3+} \\ +\mathrm{RR}_{\mathrm{X}_1+\mathrm{X}_2-\mathrm{X}_3-} + \mathrm{RR}_{\mathrm{X}_1-\mathrm{X}_2+\mathrm{X}_3-} + \mathrm{RR}_{\mathrm{X}_1-\mathrm{X}_2-\mathrm{X}_3+} \\ -\mathrm{RR}_{\mathrm{X}_1-\mathrm{X}_2-\mathrm{X}_3-} \quad (3)$$

Finally, we note that the three-way interaction (see 3) reflects the contrast of interactions between two variables over the strata of a third. We show in Appendix (Section A) that the corresponding formulae are

$$\mathrm{RERI}_3(\mathrm{X}_1, \mathrm{X}_2, \mathrm{X}_3) = \left(\mathrm{RERI}_2(\mathrm{X}_j, \mathrm{X}_k | \mathrm{X}_l = 1) * \mathrm{RR}_{\mathrm{X}_j-\mathrm{X}_k-\mathrm{X}_l+}\right) - \mathrm{RERI}_2(\mathrm{X}_j, \mathrm{X}_k | \mathrm{X}_l = 0) \quad (4)$$

because

$$\mathrm{RERI}(\mathrm{X}_j, \mathrm{X}_k | \mathrm{X}_l = 1) = \frac{\left(\mathrm{RR}_{\mathrm{X}_j+\mathrm{X}_k+\mathrm{X}_l+} - \mathrm{RR}_{\mathrm{X}_j+\mathrm{X}_k-\mathrm{X}_l+} - \mathrm{RR}_{\mathrm{X}_j-\mathrm{X}_k+\mathrm{X}_l+} + \mathrm{RR}_{\mathrm{X}_j-\mathrm{X}_k-\mathrm{X}_l+}\right)}{\mathrm{RR}_{\mathrm{X}_j-\mathrm{X}_k-\mathrm{X}_l+}}$$

given that $\mathrm{RR}_{\mathrm{X}_j-\mathrm{X}_k-\mathrm{X}_l-}$ is the reference relative risk.

where j,k,l = (1,2,3) and $j \neq k$, $j \neq l$, $k \neq l$.

The two-way interactions, given the third factor is absent ($\mathrm{RERI}_2(\mathrm{X}_1, \mathrm{X}_2 | \mathrm{X}_3 = 0)$, $\mathrm{RERI}_2(\mathrm{X}_1, \mathrm{X}_3 | \mathrm{X}_2 = 0)$ and $\mathrm{RERI}_2(\mathrm{X}_2, \mathrm{X}_3 | \mathrm{X}_1 = 0)$, as well as the corresponding interactions when the 3rd risk factor is present ($\mathrm{RERI}_2(\mathrm{X}_1, \mathrm{X}_2 | \mathrm{X}_3 = 1)$, $\mathrm{RERI}_2(\mathrm{X}_1, \mathrm{X}_3 | \mathrm{X}_2 = 1)$ and $\mathrm{RERI}_2(\mathrm{X}_2, \mathrm{X}_3 | \mathrm{X}_1 = 1)$) are important measures in the study of joint effects of 3 factors. They are very helpful in better specifying under which conditions 2 of the 3 factors interact. In the classic framework of 2-way interactions,

researchers report a specific value for RERI between two variables $X_1$ and $X_2$. However, this value may not be constant across the strata of a 3rd factor $X_3$ and to check for that issue (which was named "the uniqueness problem" by Skrondal[12]), we can calculate $\text{RERI}_2(X_1, X_2|X_3 = 0)$ and $\text{RERI}_2(X_1, X_2|X_3 = 1)$.

In the Appendix (section B), we show how to calculate the formulae of all these indexes for additive interaction in the presence of 3 risk factors (both 2 and 3-way), when applying Cox regression. The formulae are the same when using logistic regression as well. Finally, we provide user-friendly Stata code that would be useful for researchers who wish to implement these methods and calculate all possible 2- and 3-way interactions (Appendix, section B). Below we briefly present some recommendations on how to implement 3-way interaction analysis [in Appendix (Section D), we present more details on each of these steps].

1) Conduct the analysis with the exposure of interest, <u>without using any interaction term</u> to check whether the exposures are risk or protective factors. If any of the exposures appeared to be a protective factor, then recode it to a risk factor
2) Perform the analysis of the 3 risk factors, this time <u>using their interaction terms</u>
3) Compute $\text{TotRERI}_3$ and find out whether the effects of the risk factors are super- or sub-additive
4) Estimate $\text{RERI}_3$ to check whether any deviation from additivity of the three risk factors ($\text{TotRERI}_3$) is due to the 3-way interaction, beyond the two way interactions [see equation (3)].
5) Calculate the two way interactions, given the 3rd is absent [from (A.SB.4)–(A.SB.6), Appendix, Section B) to test whether any deviation from additivity of the three risk factors (expressed through $TotRERI_3$) is attributed to additive interaction of the two risk factors [(see equation (3)].

6) Compute the two way interactions, given the 3rd is present (from (A.SB.7)–(A.SB.9) in Appendix, Section B) to check to what extent the two-way interactions vary across the strata of a third variable.
7) Check whether each risk factor increase risk in all possible combinations with other risk factors, i.e. whether there is qualitative interaction (when the exposure is a risk factor for a specific outcome for one subgroup, but a protective factor for another subgroup). This information would be useful for decision making for public health purposes, because, in such instances, we should not treat all the subgroups, but only those people for which the medication is beneficial. For more details, see [16] and Appendix (Section B and F).

*From the three to the multi-way interaction on the additive scale*

When studying the multi-way interaction of n risk factors $X_1, X_2, ..., X_n$, on the additive scale, we can calculate the total relative excess risk due to interaction ($\mathrm{TotRERI_n}$), as a generalization of equation 1. $\mathrm{TotRERI_n}$ expresses the contrast of the excess risk from the situation when all risk factors act jointly versus the extra risk of the occasion that each of them acts separately, i.e.

When studying the multi-way interaction of n risk factors $X_1$, $X_2$ ..., $X_n$, on the additive scale, the comparison of interest is

$$\left(\mathrm{RR}_{X_1+X_2+\cdots X_n+} - \mathrm{RR}_{X_1-X_2-\cdots X_n-}\right) \text{ vs } \left(\mathrm{RR}_{X_1+X_2-\cdots X_n-} - \mathrm{RR}_{X_1-X_2-\cdots X_n-}\right)$$

$$+\left(RR_{X_1-X_2+\cdots X_n-} - RR_{X_1-X_2-\cdots X_n-}\right)$$

$$+\cdots$$

$$+\left(RR_{X_1-X_2-\cdots X_n+} - RR_{X_1-X_2-\cdots X_n-}\right)$$

The difference from this comparison should correspond to the total relative excess risk due to interaction, that is the sum of the n-way interaction and all the (n-1)-, (n-2)-, ..., 2-way interactions of these risk factors. In other words, we have that,

$$\mathrm{TotRERI}_n(X_1, X_2, \ldots, X_n) = \ \mathrm{ERR}_{X_1+X_2+\cdots X_n+}$$

$$-\mathrm{ERR}_{X_1+X_2-\cdots X_n-} - \mathrm{ERR}_{X_1-X_2+\cdots X_n-} - \cdots - \mathrm{ERR}_{X_1-X_2-\cdots X_n+}$$

(5)

Given that $\mathrm{TotRERI}_n$ is attributed to all potential interactions between the n variables, in other words it can be expressed as the sum of the n-way interaction and all the (n-1)-, (n-2)-, …, 2-way interactions (see Appendix, section C), i.e.

$$\mathrm{TotRERI}_n(X_1, X_2, \ldots, X_n) = \mathrm{RERI}_n(X_1, X_2, \ldots, X_n)$$

$$+ \sum_{\binom{n}{n-1}} \mathrm{RERI}_{n-1}(X_1, X_2, \ldots, X_n | 1 \text{ of the } X_i = 0)$$

$$+ \sum_{\binom{n}{n-2}} \mathrm{RERI}_{n-2}(X_1, X_2, \ldots, X_n | 2 \text{ of the } X_i = 0)$$

$$\cdots$$

$$+ \sum_{\binom{n}{2}} \mathrm{RERI}_2(X_1, X_2, \ldots, X_n | (n-2) \text{ of the } X_i = 0)$$

If we want to compute the n-way additive interaction $RERI_n$, without the contribution of all lower order interactions, then for $1 \leq i \leq n$ and $1 \leq k \leq n$, we let

$RR_{(k)}$ equal to $RR_{\mathrm{k\ of\ the\ X_i's=1,n-\ k\ of\ the\ X_i's=0}}$

we show in the Appendix (section C) that, $RERI_n(X_1, X_2, \ldots, X_n)$ for $n \geq 2$, is:

$$
\begin{aligned}
RERI_n(X_1, X_2, \ldots, X_n) = RR_{(n)} & \\
& - \sum_{\binom{n}{n-1}} RR_{(n-1)} \\
& + \sum_{\binom{n}{n-2}} RR_{(n-2)} \\
& \ldots \\
& + (-1)^n * \sum_{\binom{n}{0}} RR_{(0)}
\end{aligned}
\qquad (6)
$$

Of note that the last line can be written as $(-1)^n$, once $\binom{n}{0} = 1$ and $RR_{(0)} = 1$.

$RERI_n(X_1, X_2, \ldots, X_n)$ in (6) expresses the relative excess risk due to interaction of the n risk factors exclusively, without accounting for all the lower order additive interactions of these risk factors [i.e. extension of (4)].

Additionally, by extending equation (4) in multi-way interaction, we additionally show in the Appendix (section C) that $RERI_n$ can be written in terms of any 2 of the lower order (n-1) interactions, more specifically

$$
\begin{aligned}
& RERI_n(X_1, X_2, \ldots, X_n) \\
& \quad = \left(RERI_{n-1}(X_1, X_2, \ldots, X_n | X_i = 1) * RR_{X_1 - X_2 - \ldots X_{i+1} - X_i + X_{i+1} - \cdots X_{i+1} - \cdots X_n -}\right) \\
& \quad\quad - RERI_{n-1}(X_1, X_2, \ldots, X_n | X_i = 0)
\end{aligned}
\qquad (7)
$$

In Appendix (section D), we give the corresponding suggestions and recommendations for researchers who want to implement analysis for multi-way interactions in detail. Moreover, in Appendix (Section E), we show i) that multiplicative or super-multiplicative effects imply super-additive effects and ii) that additive or sub-additive effects imply sub-multiplicative effects for n-way interactions as well.

**Worked Example**

To illustrate the formulae derived in the previous sections we have used data from adult women participating in the Greek-EPIC study[29-30] to study the joint effects of low adherence to Mediterranean Diet (MD), obesity and smoking status on mortality. We applied survival analysis with Cox regression using as endpoint death from any cause. Levels of the indicated risk factors denoting potentially increased risk of death were i) low (scores 0-3 vs 4-9) adherence to MD, ii) obesity [Body Mass Index (BMI)≥30 kg/m$^2$ vs <30 kg/m$^2$], and (iii) smoking status (current vs. former and current) at recruitment. Age (in years) and education (4 levels; categorically modeled) were included as possible confounders. Participants with missing values in any of the above variables were excluded, leaving 15,903 women. Descriptive statistics of all variables included in the analysis are presented in table 1. In the Cox model, we included three terms for each risk factor, three terms for the 2-way product terms between those factors and one for the 3-way product term of all three factors.

The respective $TotRERI_3$, $RERI_3$ between the indicated risk factors, as well as their components i.e all 2- and 3-way interactions have been estimated using equations (A.SB.2)-(A.SB.9) (see Appendix, Section B). The Stata code that was used can be found online on github (https://github.com/mkatsoulis82/Multi-way_interaction/blob/master/Multi-way%20interaction.do ), as well as in Appendix (Section B) For the estimation of 95% confidence intervals, we used the delta method. In Table 2, we present the mortality hazard ratios, of low adherence to MD, obesity and smoking and of their joint effects as estimated by Cox regression [model (A.SB.1) in the Appendix (section B)].

**RESULTS**

From Table 2, we conclude that the effects of low MD, obesity and smoking status on mortality were super-additive ($TotRERI_3$=1.20, even not statistically significant), meaning that there was an extra 120% risk due to the joint presence of all risk factors, compared to the situation that each of them would act separately (see equation 1). More specifically, the 3-way interaction of these factors beyond the 2-way interactions was positive ($RERI_3$=1.98), indicating that there was a ~200% excess risk which is explicitly due to the 3-way interaction. On the other hand, all the $RERI_2$s given the absence of the 3$^{rd}$ risk factor are negative, even not statistically significant (first 3 rows in table 2 referring to RERIs), which is an indication that the relative risk from joint action of any two of the following: having low MD score, being a smoker and being obese, when the third factor is absent, is lower compared to sum of the relative risks of these risk factors, when acting separately (sub-additive effects). This means that the excess 120% risk due to the joint presence of all risk factors (TotRERI=1.20) is largely due to the 3-way interaction of the 3 risk factors itself ($RERI_3$=1.98), as the contribution of the 2-way interactions is negative (see equation 2). Moreover, the corresponding 2-way interactions are positive, when the third risk factor is present, which is reflected by the 3-way interaction that can be expressed in terms of the 2-way interactions [see (4)]. Finally, there was no qualitative interaction in this example.

## DISCUSSION

In this paper, we pointed out the questions of interest in the study of the joint action of >2 binary factors on a health outcome and we proposed the appropriate solutions, by introducing formulae for additive interaction. Previous publications on interactions on the additive scale refer almost exclusively to two-way interactions probably for reasons related to easiness in interpretability and communication of results.

Given this gap in the relevant studies of multi-way interactions on the additive scale, our results are novel for epidemiological research that focuses on the joint action of >2 exposures. We introduced the term "total relative excess risk due to interaction (TotRERI)", a quantity that encompasses all intermediate levels of interaction in the presence of three or more factors. Our formulae enable better interpretability of any evidence for deviations from additivity owned to more than two risk factors and provide simple ways of communicating such results from a public health perspective by attributing any excess relative risk to specific combinations of these factors. Regarding the limitations, researchers should also be concerned whether all possible categories defined by the absence and the presence of the n risk factors include sufficient number of participants, so that all RR's from expressions (3), for 3 risk factors, or (6), for n risk factors, can be adequately estimated. For case-control studies our results apply for rare diseases only, taking into consideration certain limitations that have expressed in the relevant literature, when using logistic regression[12]. Finally, the problem of the limited power in calculating 2-way interactions[17] is also present in multi-way interaction. As observed from our worked example, all the additive interactions were had very wide confidence intervals.

## CONCLUSIONS

Given the increasing interest in investigating and evaluating interactions, our results are important for studying multi-way interactions between risk factors and identifying combinations the joint presence of which may be especially important to avoid from a public health perspective.

## ABBREVIATIONS

BMI: Body Mass Index

ERR: Excess relative risk

EPIC: European Prospective Investigation into Cancer and nutrition

MD: Mediterranean Diet

RERI: Relative excess risk due to interaction

RR: Relative risk

TotRERI: Total relative excess risk due to interaction

## DECLARATIONS

### Ethics approval and consent to participate

The procedures implemented in the EPIC study were in accordance with the Declaration of Helsinki on the Ethical Principles for Medical Research involving Human Subjects of 1975 as revised in 1983. All study participants signed an informed consent form and the study protocol was approved by the ethics committees of the International Agency for Research on Cancer and the Medical School of the University of Athens.

## Consent for publication

Antonia Trichopoulou is the Principal Investigator for the Greek segment of the EPIC cohort (http://epic.iarc.fr/centers/greece.php ). The authors had Antonia Trichopoulou's permission to use data from the Greek EPIC study (worked example).

## Availability of data and material

The data that has been used is confidential. Nevertheless, the authors provided the scripts to implement multi-way interaction analysis (uploaded on github and presented in the Appendix) and provide recommendations in the paper so that the implementation of multi-way interaction analysis could be straightforward by other researchers

## Competing interests

None of the authors have any competing interest.

## Funding

Michail Katsoulis is supported by the British Heart Foundation (BHF Immediate Postdoctoral Basic Science Research Fellow – Award number **FS/18/5/33319**).

Authors' contributions

## Authors' contributions

Michail Katsoulis conceived the idea of the paper, developed the formulae, made the proofs and wrote the paper. Christina Bamia helped in the writing of the paper

**Acknowledgements**

The authors would like to thank Antonia Trichopoulou for making available EPIC-Greece data.

Table 1: Descriptive statistics of the characteristics of 15903 women participating in analysis

| CONTINUOUS VARIABLES | |
|---|---|
| | mean (sd) |
| **Age** (in years) , mean (sd) | 53.4 (12.5) |
| **CATEGORICAL VARIABLES** | |
| | N (%) |
| **BMI** | |
| Obese (BMI>=30) | 6206 (39%) |
| Non-Obese (BMI<30) | 9697 (61%) |
| **Mediterranean diet** | |
| Low adherence (0-3) | 5466 (34%) |
| Medium-High adherence (4-9) | 10437 (66%) |
| **Smoking status** | |
| Current smokers ,n(%) | 3038 (19%) |
| Never-Former smokers ,n(%) | 12865 (81%) |
| **Education** | |
| 1st level: no education (<6 years of schooling) ,n(%) | 4100 (26%) |
| 2nd level: elementary/high school (6-11 years of schooling) ,n(%) | 6502 (41%) |
| 3rd level: lyceum/technical lyceum (12 years of schooling) ,n(%) | 2872 (18%) |
| 4th level: at least university degree (>12 years of schooling) ,n(%) | 2429 (15%) |
| **Mortality** | |
| Alive till the end of follow-up | 14699 (92%) |
| Dead during follow-up | 1204 (8%) |

Table 2: Estimated hazard ratios of Mediterranean Diet (MD), obesity and smoking and of their product terms from the Cox regression (A.SB.1) from the mortality analysis conducted in women from the EPIC-Greece cohort, along with indexes of additive interaction between low adherence to MD, obesity and smoking.

## Results from Cox regression*

| Risk factors of interest and their product terms | b | se(b) | HR | 95% CI for HR |
|---|---|---|---|---|
| low MD | 0.36 | 0.09 | 1.43 | 1.20 , 1.71 |
| high BMI | 0.29 | 0.08 | 1.34 | 1.14 , 1.56 |
| never or former smokers | 0.41 | 0.18 | 1.51 | 1.05 , 2.16 |
| (low MD) * (high BMI) | -0.27 | 0.12 | 0.77 | 0.60 , 0.97 |
| (low MD) * (never or former smokers) | -0.23 | 0.30 | 0.79 | 0.44 , 1.44 |
| (high BMI) * (never or former smokers) | -0.24 | 0.29 | 0.79 | 0.45 , 1.40 |
| (low MD) * ( high BMI) * (never or former smokers) | 0.92 | 0.45 | 2.51 | 1.04 , 6.02 |

## Relative excess risk due to interaction (RERI)

| | RERI | se(RERI) | 95 % CI for RERI |
|---|---|---|---|
| $RERI_2$(low MD, high BMI / never or former smokers) | -0.30 | 0.17 | -0.64 , 0.03 |
| $RERI_2$(low MD, current smokers / low BMI) | -0.23 | 0.49 | -1.19 , 0.74 |
| $RERI_2$(high BMI, current smokers / high MD) | -0.25 | 0.45 | -1.13 , 0.63 |
| $RERI_2$(low MD, high BMI / current smokers) | 1.11 | 0.63 | -0.12 , 2.35 |
| $RERI_2$(low MD, current smokers / high BMI) | 1.31 | 0.65 | 0.05 , 2.58 |
| $RERI_2$(high BMI, current smokers / low MD) | 1.20 | 0.62 | -0.01 , 2.41 |
| $RERI_3$(low MD, high BMI ,current smokers) | 1.98 | 1.01 | 0.00 , 3.96 |
| $TotRERI_3$(low MD, high BMI ,current smokers) | 1.20 | 0.83 | -0.43 , 2.82 |

*In the Cox regression (model A.SB.1), we adjusted for age (in years) and educational levels (<6, 6-11, 12 and >12 years of schooling; categorically)

# APPENDIX

## Section A – Interaction between 3 risk factors

We have that

$$\mathrm{RERI}_2(X_1, X_2 \,|X_3 = 0) = RR_{X_1+X_2+X_3-} - RR_{X_1+X_2-X_3-} - RR_{X_1-X_2+X_3-} + RR_{X_1-X_2-X_3-} \qquad \text{(A.SA.1)}$$

$$\mathrm{RERI}_2(X_1, X_3 \,|X_2 = 0) = RR_{X_1+X_2-X_3+} - RR_{X_1+X_2-X_3-} - RR_{X_1-X_2-X_3+} + RR_{X_1-X_2-X_3-} \qquad \text{(A.SA.2)}$$

$$\mathrm{RERI}_2(X_2, X_3 \,|X_1 = 0) = RR_{X_1-X_2+X_3+} - RR_{X_1-X_2+X_3-} - RR_{X_1-X_2-X_3+} + RR_{X_1-X_2-X_3-} \qquad \text{(A.SA.3)}$$

So, by replacing in equation 2 (in the paper) to $\mathrm{TotRERI}_3(X_1, X_2, X_3)$,$\mathrm{RERI}_2(X_1, X_2 \,|X_3 = 0)$, $\mathrm{RERI}_2(X_1, X_3 \,|X_2 = 0)$ and $\mathrm{RERI}_2(X_2, X_3 \,|X_1 = 0)$ from equations (1) (in the paper), (A.SA.1), (A.SA.2), (A.SA.3) respectively, then $\mathrm{RERI}_3(X_1, X_2, X_3)$ can be also written as a function of RRs as follows:

$$\begin{aligned}\mathrm{RERI}_3(X_1, X_2, X_3) = {} & RR_{X_1+X_2+X_3+} \\ & -RR_{X_1+X_2+X_3-} - RR_{X_1+X_2-X_3+} - RR_{X_1-X_2+X_3+} \\ & +RR_{X_1+X_2-X_3-} + RR_{X_1-X_2+X_3-} + RR_{X_1-X_2-X_3+} \\ & -RR_{X_1-X_2-X_3-}\end{aligned}$$

as presented in expression 3 (in the paper)

Moreover, we have that

$$\mathrm{RERI}_2(X_1, X_2|X_3 = 1) = \frac{(RR_{X_1+X_2+X_3+} - RR_{X_1+X_2-X_3+} - RR_{X_1-X_2+X_3+} + RR_{X_1-X_2-X_3+})}{RR_{X_1-X_2-X_3+}} \qquad \text{(A.SA.4)}$$

$$\mathrm{RERI}_2(X_1, X_3|X_2 = 1) = \frac{(\mathrm{RR}_{X_1+X_2+X_3+} - \mathrm{RR}_{X_1+X_2+X_3-} - \mathrm{RR}_{X_1-X_2+X_3+} + \mathrm{RR}_{X_1-X_2+X_3-})}{\mathrm{RR}_{X_1-X_2+X_3-}} \quad \text{(A.SA.5)}$$

$$\mathrm{RERI}_2(X_2, X_3|X_1 = 1) = \frac{(\mathrm{RR}_{X_1+X_2+X_3+} - \mathrm{RR}_{X_1+X_2+X_3-} - \mathrm{RR}_{X_1+X_2-X_3+} + \mathrm{RR}_{X_1+X_2-X_3-})}{\mathrm{RR}_{X_1+X_2-X_3-}} \quad \text{(A.SA.6)}$$

Of note, we standardize $\mathrm{RERI}_2$ when the 3rd variable is present by the RR of the 3rd variable, because in the analysis we conduct, all RR's are calculated taking into consideration that $\mathrm{RR}_{X_1-X_2-X_3-}$ is the reference relative risk. However, in (A.SA.4), we are interested in

- $\mathrm{RR}_{X_1+X_2+}$, given that $X_3$ is present, compared to $\mathrm{RR}_{X_1-X_2-}$, given that $X_3$ is present, so the relative risk of interest will be $\mathrm{RR}_{X_1+X_2+X_3+}/\mathrm{RR}_{X_1-X_2-X_3+}$
- $\mathrm{RR}_{X_1+X_2-}$, given that $X_3$ is present, compared to $\mathrm{RR}_{X_1-X_2-}$, given that $X_3$ is present, so the relative risk of interest will be $\mathrm{RR}_{X_1+X_2-X_3+}/\mathrm{RR}_{X_1-X_2-X_3+}$
- $\mathrm{RR}_{X_1-X_2+}$, given that $X_3$ is present, compared to $\mathrm{RR}_{X_1-X_2-}$, given that $X_3$ is present, so the relative risk of interest will be $\mathrm{RR}_{X_1-X_2+X_3+}/\mathrm{RR}_{X_1-X_2-X_3+}$
- $\mathrm{RR}_{X_1-X_2-}$, given that $X_3$ is present, compared to $\mathrm{RR}_{X_1-X_2-}$, given that $X_3$ is present, so the relative risk of interest will be $\mathrm{RR}_{X_1-X_2-X_3+}/\mathrm{RR}_{X_1-X_2-X_3+}$

By combining these bullets, it is straightforward why we standardize by $\mathrm{RR}_{X_1-X_2-X_3+}$ in (A.SA.4). For the same reasons we standardized by $\mathrm{RR}_{X_1-X_2+X_3-}$ and $\mathrm{RR}_{X_1+X_2-X_3-}$ in (A.SA.5) and (A.SA.6).

Moreover, it is also straightforward that if we combine (A.SA.1) and (A.SA.4), or (A.SA.2) and (A.SA.5), or (A.SA.3) and (A.SA.6), we end up to equation (4) in the manuscript.

## Section B – Estimation of interactions between 3 risk factors when using Cox (or logistic) regression and implementation in STATA

### *Estimations using Cox regression*

We consider dichotomous variables $X_1$, $X_2$ and $X_3$ as risk factors for the outcome of interest (disease D in incidence analysis or death in mortality analysis) with $X_i$=(0,1) referring to the absence/presence of risk factor $X_i$ hypothesized to be associated with the lowest/highest risk for disease D). Let also $U_j$, j=1,2,...,n denote additional variables which serve as potential confounders to the association of $X_i$ with D. Finally, let the hazard rate for disease D at time t be $\lambda(t) = \lambda(t, X_1, X_2, X_3, Z_j)$ as estimated from Cox regression model including $X_1$, $X_2$ and $X_3$. and additional covariates $X_1X_2$,$X_1X_3$, $X_2X_3$ and $X_1X_2X_3$ for all 2 and 3 way interactions of $X_1$, $X_2$ and $X_3$, and controlling for n covariates $U_j$, j=1,2,...,n :

$$\lambda(t) = \lambda_0(t) * \exp(a_1X_1 + a_2X_2 + a_3X_3 + a_4X_1X_2 + a_5X_1X_3 + a_6X_2X_3 + a_7X_1X_2X_3 + \textstyle\sum_{j=1}^{n} c_jU_j), \quad \text{(A.SB.1)}$$

where $a_k$, k=1,..,7 and $c_j$, j=1,..,N are the log of the hazard ratios estimated from the Cox model. Based on model (A.SB.1), $TotRERI_3(X_1,X_2,X_3)$ and $RERI_3$ $(X_1, X_2, X_3)$ can be estimated as shown below:

$$\mathrm{TotRERI}_3(X_1, X_2, X_3) = RR_{X_1+X_2+X_3+} - RR_{X_1+X_2-X_3-} - RR_{X_1-X_2+X_3-} - RR_{X_1-X_2-X_3+} + 2$$

$$= \exp(a_1 + a_2 + a_3 + a_4 + a_5 + a_6 + a_7) - \exp(a_1) - \exp(a_2) - \exp(a_3) + 2 \quad \text{(A.SB.2)}$$

and

$$\mathrm{RERI}_3(X_1, X_2, X_3) = RR_{X_1+X_2+X_3+} - RR_{X_1+X_2+X_3-} - RR_{X_1+X_2-X_3+} - RR_{X_1-X_2+X_3+}$$

$$+RR_{X_1+X_2-X_3-} + RR_{X_1-X_2+X_3-} + RR_{X_1-X_2-X_3+} - RR_{X_1-X_2-X_3-}$$

$$= \exp(a_1 + a_2 + a_3 + a_4 + a_5 + a_6 + a_7) - \exp(a_1 + a_2 + a_4) - \exp(a_1 + a_3 + a_5)$$

$$-\exp(a_2 + a_3 + a_6) + \exp(a_1) + \exp(a_2) + \exp(a_3) - 1 \qquad \text{(A.SB.3)}$$

Similarly, one can estimate also the components of $TotRERI_3(X_1,X_2,X_3)$ and $RERI_3(X_1, X_2, X_3)$, based on (A.SB.1):

$$RERI_2(X_1, X_2|X_3 = 0) = \exp(a_1 + a_2 + a_4) - \exp(a_1) - \exp(a_2) + 1 \qquad \text{(A.SB.4)}$$

$$RERI_2(X_1, X_3|X_2 = 0) = \exp(a_1 + a_3 + a_5) - \exp(a_1) - \exp(a_3) + 1 \qquad \text{(A.SB.5)}$$

$$RERI_2(X_2, X_3|X_1 = 0) = \exp(a_2 + a_3 + a_6) - \exp(a_2) - \exp(a_3) + 1 \qquad \text{(A.SB.6)}$$

$$RERI_2(X_1, X_2|X_3 = 1) = \frac{\exp(a_1+a_2+a_3+a_4+a_5+a_6+a_7) - \exp(a_1+a_3+a_5) - \exp(a_2+a_3+a_6) + \exp(a_3)}{\exp(a_3)} \qquad \text{(A.SB.7)}$$

$$RERI_2(X_1, X_3|X_2 = 1) = \frac{\exp(a_1+a_2+a_3+a_4+a_5+a_6+a_7) - \exp(a_1+a_2+a_4) - \exp(a_2+a_3+a_6) + \exp(a_2)}{\exp(a_2)} \qquad \text{(A.SB.8)}$$

$$RERI_2(X_2, X_3|X_1 = 1) = \frac{\exp(a_1+a_2+a_3+a_4+a_5+a_6+a_7) - \exp(a_1+a_2+a_4) - \exp(a_1+a_3+a_5) + \exp(a_1)}{\exp(a_1)} \qquad \text{(A.SB.9)}$$

***Implementation in STATA***

```
/*

DESCRIPTION

time; survival time

death (outcome); 0-->alive, 1-->dead

x1 (non adherence to Mediterranean diet - 1st risk factor);

0--> high adherence to Mediterranean diet, 1--> low adherence to Mediterranean diet


x2 (being obese - 2nd risk factor);

0--> not being obese (BMI<30), 1--> being obese (BMI>=30)


x3 (smoking status - 3rd risk factor);

0--> never/former smoker, 1--> current smoker
```

u1 (age - 1st confounder); continuous in years

u2 (education level - 2nd confounder); categorical in 4 levels

*/

```
* IMPLEMENTATION

* At first we compute the product terms

gen x1x2=x1*x2

gen x1x3=x1*x3

gen x2x3=x2*x3

gen x1x2x3=x1*x2*x3


*We run the Cox model

stset time, failure(death)


stcox x1 x2 x3 x1x2 x1x3 x2x3 x1x2x3 u1 i.u2


*We compute TotRERI3

nlcom TotRERI3: exp(_b[x1]+_b[x2]+_b[x3]+_b[x1x2]+_b[x1x3]+_b[x2x3]+_b[x1x2x3])-exp(_b[x1])-exp(_b[x2])-exp(_b[x3])+2


*We compute RERI3

nlcom RERI3: exp(_b[x1]+_b[x2]+_b[x3]+_b[x1x2]+_b[x1x3]+_b[x2x3]+_b[x1x2x3])-exp(_b[x1]+_b[x2]+_b[x1x2])-exp(_b[x1]+_b[x3]+_b[x1x3])-exp(_b[x2]+_b[x3]+_b[x2x3])+exp(_b[x1])+exp(_b[x2])+exp(_b[x3])-1
```

*We compute 2-way interactions, given the 3rd risk factor is absent

*RERI(x1,x2/x3=0)

nlcom RERI2_x1_x2_given_x3is0: exp(_b[x1]+_b[x2]+_b[x1x2])-exp(_b[x1])-exp(_b[x2])+1

*RERI(x1,x3/x2=0)

nlcom RERI2_x1_x3_given_x2is0: exp(_b[x1]+_b[x3]+_b[x1x3])-exp(_b[x1])-exp(_b[x3])+1

*RERI(x2,x3/x1=0)

nlcom RERI2_x2_x3_given_x1is0: exp(_b[x2]+_b[x3]+_b[x2x3])-exp(_b[x2])-exp(_b[x3])+1

*We compute 2-way interactions, given the 3rd risk factor is present

*RERI(x1,x2/x3=1)

nlcom RERI2_x1_x2_given_x3is1:
(exp(_b[x1]+_b[x2]+_b[x3]+_b[x1x2]+_b[x1x3]+_b[x2x3]+_b[x1x2x3])-
exp(_b[x1]+_b[x3]+_b[x1x3])-exp(_b[x2]+_b[x3]+_b[x2x3])+exp(_b[x3]))/exp(_b[x3])

*RERI(x1,x3/x2=1)

nlcom RERI2_x1_x3_given_x2is1:
(exp(_b[x1]+_b[x2]+_b[x3]+_b[x1x2]+_b[x1x3]+_b[x2x3]+_b[x1x2x3])-
exp(_b[x1]+_b[x2]+_b[x1x2])-exp(_b[x2]+_b[x3]+_b[x2x3])+exp(_b[x2]))/exp(_b[x2])

*RERI(x2,x3/x1=1)

nlcom RERI2_x2_x3_given_x1is1:
(exp(_b[x1]+_b[x2]+_b[x3]+_b[x1x2]+_b[x1x3]+_b[x2x3]+_b[x1x2x3])-
exp(_b[x1]+_b[x2]+_b[x1x2])-exp(_b[x1]+_b[x3]+_b[x1x3])+exp(_b[x1]))/exp(_b[x1])

*The same formulae for all these RERIs are used when running logistic regression

* CHECK FOR QUALITATIVE INTERACTION

*We run again the Cox model

stset time, failure(death)

stcox x1 x2 x3 x1x2 x1x3 x2x3 x1x2x3 u1 i.u2

* To check whether the risk for x1 is increasing across strata of x2,x3, we have to examine whether the following quantities are positive (i.e. >0)

* 1a) to see if RR100>RR000, we check whether RR100-RR000>0

disp exp(_b[x1])-1

* 1b) to see if RR110>RR010, we check whether RR110-RR010>0

disp exp(_b[x1]+_b[x2]+_b[x1x2])-exp(_b[x2])

* 1c) to see if RR101>RR001, we check whether RR101-RR001>0

disp exp(_b[x1]+_b[x3]+_b[x1x3])-exp(_b[x3])

* 1d) to see if RR111>RR011, we check whether RR111-RR011>0

disp exp(_b[x1]+_b[x2]+_b[x3]+_b[x1x2]+_b[x1x3]+_b[x2x3]+_b[x1x2x3])-exp(_b[x2]+_b[x3]+_b[x2x3])

* To check whether the risk for x2 is increasing across strata of x1,x3

* 2a) to see if RR010>RR000, we check whether RR010-RR000>0

disp exp(_b[x2])-1

* 2b) to see if RR110>RR100, we check whether RR110-RR100>0

disp exp(_b[x1]+_b[x2]+_b[x1x2])-exp(_b[x1])

* 2c) to see if RR011>RR001, we check whether RR011-RR001>0

disp exp(_b[x2]+_b[x3]+_b[x2x3])-exp(_b[x3])

* 2d) to see if RR111>RR101, we check whether RR111-RR101>0

disp exp(_b[x1]+_b[x2]+_b[x3]+_b[x1x2]+_b[x1x3]+_b[x2x3]+_b[x1x2x3])-exp(_b[x1]+_b[x3]+_b[x1x3])

* To check whether the risk for x3 is increasing across strata of x1,x2

* 2a) to see if RR001>RR000, we check whether RR001-RR000>0

disp exp(_b[x3])-1

* 2b) to see if RR101>RR100, we check whether RR101-RR010>0

disp exp(_b[x1]+_b[x3]+_b[x1x3])-exp(_b[x1])

* 2c) to see if RR011>RR010, we check whether RR011-RR100>0

disp exp(_b[x2]+_b[x3]+_b[x2x3])-exp(_b[x2])

* 2d) to see if RR111>RR110, we check whether RR111-RR110>0

disp exp(_b[x1]+_b[x2]+_b[x3]+_b[x1x2]+_b[x1x3]+_b[x2x3]+_b[x1x2x3])-exp(_b[x1]+_b[x2]+_b[x1x2])

**Section C – proofs of equations of multi way interaction**

*Proof of expression (6)*

We define the excess relative risk by $ERR_{X_1\#X_2\#\ldots X_n\#}$ , where # = +/-, as

$ERR_{X_1\#X_2\#\ldots X_n\#} = RR_{X_1\#X_2\#\ldots X_n\#} - RR_{X_1-X_2-\cdots X_n-}$ , i.e.

$ERR_{X_1\#X_2\#\ldots X_n\#} = RR_{X_1\#X_2\#\ldots X_n\#} - 1$

Now, expression (5) from the text can be written as

$$TotRERI_n(X_1, X_2, \ldots, X_n) = ERR_{X_1+X_2+\cdots X_n+}$$

$$-ERR_{X_1+X_2-\cdots X_n-} - ERR_{X_1-X_2+\cdots X_n-} - \cdots - ERR_{X_1-X_2-\cdots X_n+} \qquad \text{(A.SC.1)}$$

$TotRERI_n$ can be expressed in terms of relative risks (see equation 5 in the paper), but also as a function of all interactions, more specifically all $RERI_k$ for k≤n, that is

$$TotRERI_n(X_1, X_2, \ldots, X_n) = RERI_n(X_1, X_2, \ldots, X_n)$$

$$+ \sum_{\binom{n}{n-1}} RERI_{n-1}(X_1, X_2, \ldots, X_n | 1 \text{ of the } X_i = 0)$$

$$+ \sum_{\binom{n}{n-2}} RERI_{n-2}(X_1, X_2, \ldots, X_n | 2 \text{ of the } X_i = 0)$$

$$\cdots$$

$$+ \sum_{\binom{n}{2}} RERI_2(X_1, X_2, \dots, X_n | (n-2) \text{ of the } X_i = 0) \qquad (A.SC.2)$$

Now, to calculate $RERI_n$, we can combine equations (A.SC.1) and (A.SC.2), we have to solve the recurrence relation

$$RERI_n + \sum_{\binom{n}{n-1}} RERI_{n-1} + \sum_{\binom{n}{n-2}} RERI_{n-2} + \cdots + \sum_{\binom{n}{2}} RERI_2 = ERR_{X_1+X_2+\cdots X_n+} - ERR_{X_1+X_2-\cdots X_n-} - ERR_{X_1-X_2+\cdots X_n-} - \cdots - ERR_{X_1-X_2-\cdots X_n+}$$

Where $RERI_{n-k} = RERI_k(X_1, X_2, \dots, X_n | (n-k) \text{ of the } X_i = 0)$

In other words, we have to solve

$$RERI_n = ERR_{X_1+X_2+\cdots X_n+} - ERR_{X_1+X_2-\cdots X_n-} - ERR_{X_1-X_2+\cdots X_n-} - \cdots - ERR_{X_1-X_2-\cdots X_n+} - \sum_{\binom{n}{1}} RERI_{n-1} - \sum_{\binom{n}{2}} RERI_{n-2} - \cdots - \sum_{\binom{n}{n-2}} RERI_2$$

using as notation $RR_{(k)} = RR_{k \text{ of the n } X_i's=1, \text{the rest } (n-k) \ X_i's=0}$ we have that

$$RERI_n = RR_{(n)} - RR_{(0)} - \sum_{\binom{n}{1}} ERR_{(1)} - \sum_{\binom{n}{n-1}} RERI_{n-1} - \sum_{\binom{n}{n-2}} RERI_{n-2} - \cdots - \sum_{\binom{n}{2}} RERI_2$$

in other words

$$RERI_n = RR_{(n)} - \sum_{\binom{n}{n-1}} RERI_{n-1} - \sum_{\binom{n}{n-2}} RERI_{n-2} - \cdots - \sum_{\binom{n}{2}} RERI_2 - \sum_{\binom{n}{1}} ERR_{(1)} - \sum_{\binom{n}{0}} RR_{(0)} \qquad (A.SC.3)$$

when we express all the (different) $RERI_{n-1}$'s from the recurrence relation (A.SC.3), we have that

$$\mathrm{RERI}_n = \mathrm{RR}_{(n)} - \sum_{\binom{n}{n-1}} \mathrm{RR}_{(n-1)} + \sum_{\binom{n}{n-1}} \sum_{\binom{n-1}{n-2}} \mathrm{RERI}_{n-2} + \sum_{\binom{n}{n-1}} \sum_{\binom{n-1}{n-3}} \mathrm{RERI}_{n-3} + \cdots + \sum_{\binom{n}{n-1}} \sum_{\binom{n-1}{n-k}} \mathrm{RERI}_{n-k} + \cdots + \sum_{\binom{n}{n-1}} \sum_{\binom{n-1}{1}} \mathrm{ERR}_1 + \sum_{\binom{n}{n-1}} \sum_{\binom{n-1}{0}} \mathrm{RR}_0$$

$$- \sum_{\binom{n}{n-2}} \mathrm{RERI}_{n-2} - \sum_{\binom{n}{n-3}} \mathrm{RERI}_{n-3} - \ldots - \sum_{\binom{n}{k}} \mathrm{RERI}_{n-k} - \ldots - \sum_{\binom{n}{1}} \mathrm{ERR}_1 - \sum_{\binom{n}{0}} \mathrm{RR}_0 \qquad (\mathrm{A.SC.4})$$

and when we express all $\mathrm{RERI}_{n-2}$'s from (A.SC.4) using the recurrence relation (A.SC.3), we have that

$$\mathrm{RERI}_n = \mathrm{RR}_{(n)} - \sum_{\binom{n}{n-1}} \mathrm{RR}_{(n-1)} + \left( \sum_{\binom{n}{n-1}} \sum_{\binom{n-1}{n-2}} \mathrm{RR}_{(n-2)} - \sum_{\binom{n}{n-2}} \mathrm{RR}_{(n-2)} \right)$$

$$- \sum_{\binom{n}{n-1}} \sum_{\binom{n-1}{n-2}} \sum_{\binom{n-2}{n-3}} \mathrm{RERI}_{n-3} - \sum_{\binom{n}{n-1}} \sum_{\binom{n-1}{n-2}} \sum_{\binom{n-2}{n-4}} \mathrm{RERI}_{n-4} - \cdots - \sum_{\binom{n}{n-1}} \sum_{\binom{n-1}{n-2}} \sum_{\binom{n-2}{n-k}} \mathrm{RERI}_{n-k} - \cdots - \sum_{\binom{n}{n-1}} \sum_{\binom{n-1}{n-2}} \sum_{\binom{n-2}{0}} \mathrm{RR}_0$$

$$+ \sum_{\binom{n}{n-1}} \sum_{\binom{n-1}{n-3}} \mathrm{RERI}_{n-3} + \sum_{\binom{n}{n-1}} \sum_{\binom{n-1}{n-4}} \mathrm{RERI}_{n-4} + \ldots + \sum_{\binom{n}{n-1}} \sum_{\binom{n-1}{n-k}} \mathrm{RERI}_{n-k} + \ldots + \sum_{\binom{n}{n-1}} \sum_{\binom{n-1}{0}} \mathrm{RR}_0$$

$$+ \sum_{\binom{n}{n-2}} \sum_{\binom{n-2}{n-3}} \mathrm{RERI}_{n-3} + \sum_{\binom{n}{n-2}} \sum_{\binom{n-2}{n-4}} \mathrm{RERI}_{n-4} + \ldots + \sum_{\binom{n}{n-2}} \sum_{\binom{n-2}{n-k}} \mathrm{RERI}_{n-k} + \ldots + \sum_{\binom{n}{n-2}} \sum_{\binom{n-2}{0}} \mathrm{RR}_0$$

$$- \sum_{\binom{n}{n-3}} \mathrm{RERI}_{n-3} - \sum_{\binom{n}{n-4}} \mathrm{RERI}_{n-4} - \ldots - \sum_{\binom{n}{k}} \mathrm{RERI}_{n-k} - \ldots - \sum_{\binom{n}{0}} \mathrm{RR}_0 \qquad (\mathrm{A.SC.5})$$

of interest from (A.SC.5) to compute the quantity

$$\sum_{\binom{n}{n-1}} \sum_{\binom{n-1}{n-2}} RR_{(n-2)} - \sum_{\binom{n}{n-2}} RR_{(n-2)} \qquad (A.SC.6)$$

i.e. how many time we are going to sum up $RR_{(n-2)}$

from combinatorics, it is known that

$$\binom{n}{r}\binom{r}{k} = \binom{n}{k}\binom{n-k}{r-k}$$

so from (A.SC.6) we have that

$$\binom{n}{n-1}\binom{n-1}{n-2} = \binom{n}{n-2}\binom{n-n+2}{n-1-n+2} = \binom{n}{n-2}\binom{2}{1} = 2\binom{n}{n-2}$$

in other words, $\binom{n}{n-1}\binom{n-1}{n-2} - \binom{n}{n-2} = 2\binom{n}{n-2} - \binom{n}{n-2} = \binom{n}{n-2}$

so it seems that

$$\sum_{\binom{n}{n-1}} \sum_{\binom{n-1}{n-2}} RR_{(n-2)} - \sum_{\binom{n}{n-2}} RR_{(n-2)} = \sum_{\binom{n}{n-2}} RR_{(n-2)} \qquad (A.SC.7)$$

However, to prove (A.SC.7) and more specifically that the summation of $RR_{(n-2)}$ will be the sum of all different $RR_{(n-2)}$'s, we work as follows;

It is obvious that the part of the summation of $RR_{(n-2)}$ , which is created through the route

$RERI_n \rightarrow RERI_{n-2}$ is equal to

$$\sum_{\binom{n}{n-2}} RR_{(n-2)}$$

more specifically it is equal the sum of all different $RR_{(n-2)}$'s

Now, we have to prove that the part of the summation of $RR_{(n-2)}$ , which is created through the route

$RERI_n \to RERI_{n-1} \to RERI_{n-2}$, which is

$$\sum_{\binom{n}{n-1}} \sum_{\binom{n-1}{n-2}} RR_{(n-2)}$$

can be written

$$\sum_{\binom{n}{n-1}} \sum_{\binom{n-1}{n-2}} RR_{(n-2)} = 2 * \sum_{\binom{n}{n-2}} RR_{(n-2)}$$

In other words, that is equal to the sum of the $\binom{n}{n-2}$ different $RERI_{n-2}$, multiplied by 2

To show that, we work as follows;

We have $\binom{n}{n-1} = n$ different $RERI_{n-1}$'s that lead to $\binom{n}{n-1}\binom{n-1}{n-2} = 2 * \binom{n}{n-2} = 2 * \frac{(n-1)(n-2)}{2}$ $RERI_{n-2}$'s

i.e. $(n-1)(n-2)$ $RERI_{n-2}$'s through the following pattern

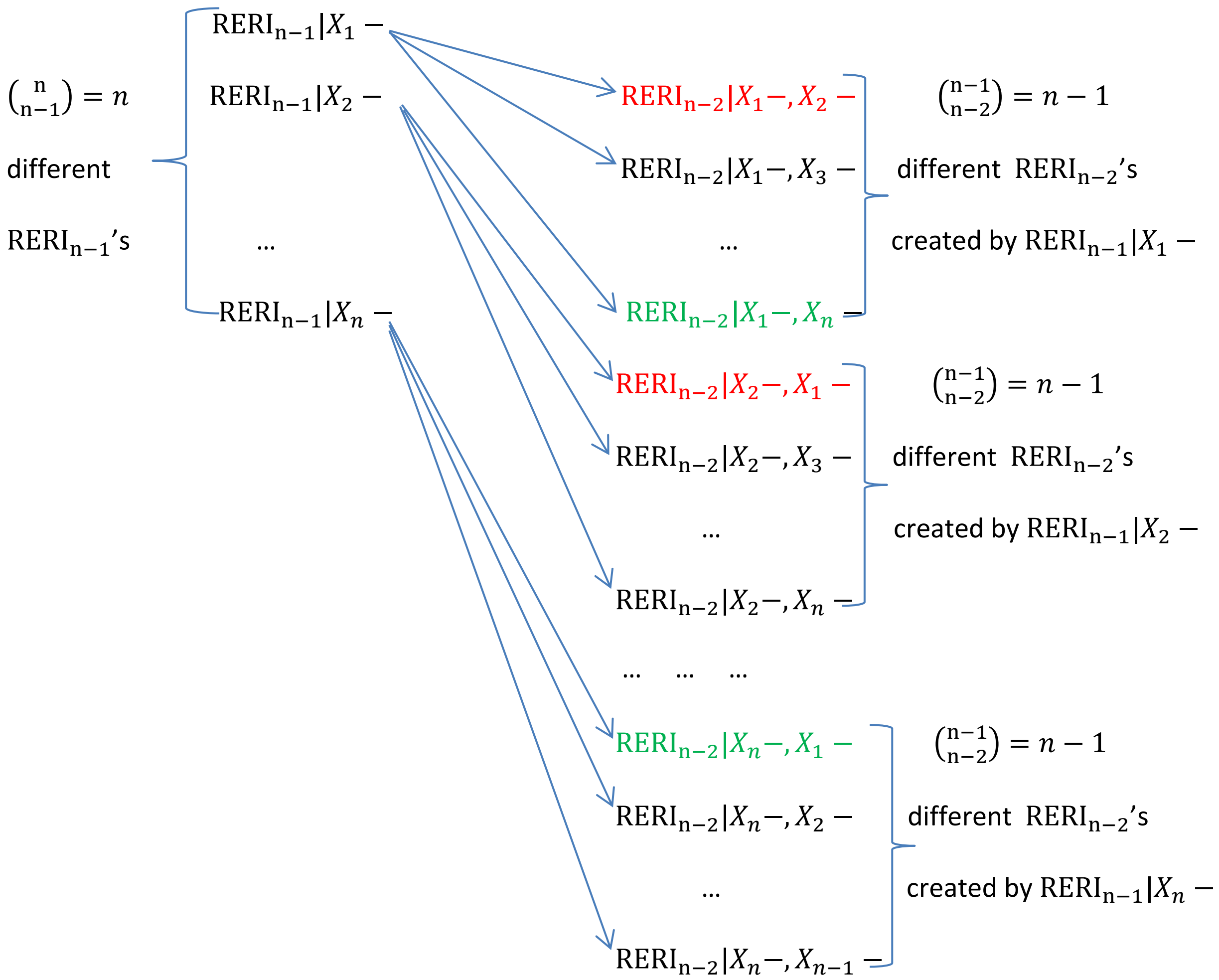

(n n−1) = n
different
RERI_{n−1}'s
RERI_{n−1}|X_1 −
RERI_{n−1}|X_2 −
…
RERI_{n−1}|X_n −
RERI_{n−2}|X_1−, X_2 −
RERI_{n−2}|X_1−, X_3 −
…
RERI_{n−2}|X_1−, X_n −
(n−1 n−2) = n − 1
different RERI_{n−2}'s
created by RERI_{n−1}|X_1 −
RERI_{n−2}|X_2−, X_1 −
RERI_{n−2}|X_2−, X_3 −
…
RERI_{n−2}|X_2−, X_n −
(n−1 n−2) = n − 1
different RERI_{n−2}'s
created by RERI_{n−1}|X_2 −
… … …
RERI_{n−2}|X_n−, X_1 −
RERI_{n−2}|X_n−, X_2 −
…
RERI_{n−2}|X_n−, X_{n−1} −
(n−1 n−2) = n − 1
different RERI_{n−2}'s
created by RERI_{n−1}|X_n −

From above, we can observe that the $\binom{\mathrm{n}}{\mathrm{n-2}}$ different $\mathrm{RERI_{n-2}}$ are created from the $\binom{\mathrm{n}}{\mathrm{n-1}} = n$ different $\mathrm{RERI_{n-1}}$'s twice. This is done because $\mathrm{RERI_{n-2}}|X_m -, X_l -$ is derived by both $\mathrm{RERI_{n-1}}|X_m -$ and $\mathrm{RERI_{n-1}}|X_l -$, but from no other $\mathrm{RERI_{n-1}}|X_j -$, j≠m, j≠l (for example see $\mathrm{RERI_{n-2}}|X_1 -, X_2 -$ and $\mathrm{RERI_{n-2}}|X_1 -, X_n -$). So the total $\mathrm{RR_{(n-2)}}$ that will be created through the route $\mathrm{RERI_n} \rightarrow \mathrm{RERI_{n-1}} \rightarrow \mathrm{RERI_{n-2}}$ is equal to

$$\sum_{\binom{\mathrm{n}}{\mathrm{n-1}}} \sum_{\binom{\mathrm{n-1}}{\mathrm{n-2}}} \mathrm{RR_{(n-2)}} = 2 * \sum_{\binom{\mathrm{n}}{\mathrm{n-2}}} \mathrm{RR_{(n-2)}}$$

After this result, (A.SC.7) is proved, because

$$\sum_{\binom{\mathrm{n}}{\mathrm{n-1}}} \sum_{\binom{\mathrm{n-1}}{\mathrm{n-2}}} \mathrm{RR_{(n-2)}} - \sum_{\binom{\mathrm{n}}{\mathrm{n-2}}} \mathrm{RR_{(n-2)}} = 2 * \sum_{\binom{\mathrm{n}}{\mathrm{n-2}}} \mathrm{RR_{(n-2)}} - \sum_{\binom{\mathrm{n}}{\mathrm{n-2}}} \mathrm{RR_{(n-2)}} = \sum_{\binom{\mathrm{n}}{\mathrm{n-2}}} \mathrm{RR_{(n-2)}}$$

Furthermore, when we express $\mathrm{RERI_{n-3}}$ from (A.SC.5) using the recurrence relation (A.SC.3), we have that

$$\mathrm{RERI_n} = \mathrm{RR_{(n)}} - \sum_{\binom{\mathrm{n}}{\mathrm{n-1}}} \mathrm{RR_{(n-1)}} + \sum_{\binom{\mathrm{n}}{\mathrm{n-2}}} \mathrm{RR_{(n-2)}} - \left( \sum_{\binom{\mathrm{n}}{\mathrm{n-1}}} \sum_{\binom{\mathrm{n-1}}{\mathrm{n-2}}} \sum_{\binom{\mathrm{n-2}}{\mathrm{n-3}}} \mathrm{RR_{(n-3)}} - \sum_{\binom{\mathrm{n}}{\mathrm{n-1}}} \sum_{\binom{\mathrm{n-1}}{\mathrm{n-3}}} \mathrm{RR_{(n-3)}} - \sum_{\binom{\mathrm{n}}{\mathrm{n-2}}} \sum_{\binom{\mathrm{n-2}}{\mathrm{n-3}}} \mathrm{RR_{(n-3)}} + \sum_{\binom{\mathrm{n}}{\mathrm{n-3}}} \mathrm{RR_{(n-3)}} \right)$$

$$+ \sum_{\binom{\mathrm{n}}{\mathrm{n-1}}} \sum_{\binom{\mathrm{n-1}}{\mathrm{n-2}}} \sum_{\binom{\mathrm{n-2}}{\mathrm{n-3}}} \sum_{\binom{\mathrm{n-3}}{\mathrm{n-4}}} \mathrm{RERI_{n-4}} + \sum_{\binom{\mathrm{n}}{\mathrm{n-1}}} \sum_{\binom{\mathrm{n-1}}{\mathrm{n-2}}} \sum_{\binom{\mathrm{n-2}}{\mathrm{n-3}}} \sum_{\binom{\mathrm{n-3}}{\mathrm{n-5}}} \mathrm{RERI_{n-5}} + \cdots + \sum_{\binom{\mathrm{n}}{\mathrm{n-1}}} \sum_{\binom{\mathrm{n-1}}{\mathrm{n-2}}} \sum_{\binom{\mathrm{n-2}}{\mathrm{n-3}}} \sum_{\binom{\mathrm{n-3}}{\mathrm{n-k}}} \mathrm{RERI_{n-k}} + \cdots + \sum_{\binom{\mathrm{n}}{\mathrm{n-1}}} \sum_{\binom{\mathrm{n-1}}{\mathrm{n-2}}} \sum_{\binom{\mathrm{n-2}}{\mathrm{n-3}}} \sum_{\binom{\mathrm{n-3}}{0}} \mathrm{RR_0}$$

$$\ldots \qquad \ldots \qquad \ldots \qquad \ldots \qquad \ldots$$

$$\cdots \qquad \cdots \qquad \cdots \qquad \cdots \qquad \cdots$$

$$\cdots \qquad \cdots \qquad \cdots \qquad \cdots \qquad \cdots$$

$$-\sum_{\binom{n}{n-4}} \mathrm{RERI}_{n-4} \quad - \quad \sum_{\binom{n}{n-5}} \mathrm{RERI}_{n-5} - \quad \ldots \quad -\sum_{\binom{n}{k}} \mathrm{RERI}_{n-k} \quad -\sum_{\binom{n}{0}} \mathrm{RR}_0 \qquad (\mathrm{A.SC.8})$$

of interest from (A.SC.8) to compute how many time we are going to sum up $\mathrm{RR}_{(n-3)}$, i.e.

$$\sum_{\binom{n}{n-1}} \sum_{\binom{n-1}{n-2}} \sum_{\binom{n-2}{n-3}} \mathrm{RR}_{(n-3)} - \sum_{\binom{n}{n-1}} \sum_{\binom{n-1}{n-3}} \mathrm{RR}_{(n-3)} - \sum_{\binom{n}{n-2}} \sum_{\binom{n-2}{n-3}} \mathrm{RR}_{(n-3)} + \sum_{\binom{n}{n-3}} \mathrm{RR}_{(n-3)} \qquad (\mathrm{A.SC.9})$$

We have that

$$\binom{n}{n-1}\binom{n-1}{n-2}\binom{n-2}{n-3} = 6\binom{n}{n-3}, \qquad \binom{n}{n-1}\binom{n-1}{n-3} = 3\binom{n}{n-3}, \qquad \binom{n}{n-2}\binom{n-2}{n-3} = 3\binom{n}{n-3}, so$$

$$6\binom{n}{n-3} - 3\binom{n}{n-3} - 3\binom{n}{n-3} + \binom{n}{n-3} = \binom{n}{n-3}$$

so if we work as we did to show (A.SC.7), we will find that

$$\left(\sum_{\binom{n}{n-1}}\sum_{\binom{n-1}{n-2}}\sum_{\binom{n-2}{n-3}} RR_{(n-3)} - \sum_{\binom{n}{n-1}}\sum_{\binom{n-1}{n-3}} RR_{(n-3)} - \sum_{\binom{n}{n-2}}\sum_{\binom{n-2}{n-3}} RR_{(n-3)} + \sum_{\binom{n}{n-3}} RR_{(n-3)}\right) = \sum_{\binom{n}{n-3}} RR_{(n-3)}$$

In other words, (A.SC.9) is equal to the sum of the $\binom{n}{n-3}$ different $RR_{(n-3)}$, that come from the application of (A.SC.3) for the conversion of $\binom{n}{n-3}$ different $RERI_{(n-3)}$'s.

Moreover, for the parenthesis related to the summation of $RR_{(n-3)}$ (i.e. (A.SC.9) we may also calculate it as follows:

We let

$\sum_{\binom{n}{n-1}}\sum_{\binom{n-1}{n-2}}\sum_{\binom{n-2}{n-3}} RR_{(n-3)} \rightarrow (1\ \ 2)_{RR_{(n-3)}}$, i.e. the part of the summation that $RR_{(n-3)}$ is calculated from $RERI_{n-3}$through the route

$RERI_n \rightarrow RERI_{n-1} \rightarrow RERI_{n-2} \rightarrow RERI_{n-3}$ (we name points [1] and [2]),

i.e. we go from $RERI_n$to $RERI_{n-3}$through $RERI_{n-1}$to $RERI_{n-2}$ (points [1] and [2] respectively)

$\sum_{\binom{n}{n-1}}\sum_{\binom{n-1}{n-3}} RR_{(n-3)} \rightarrow (1)_{RR_{(n-3)}}$, i.e. the part of the summation that $RR_{(n-3)}$ is calculated from $RERI_{n-3}$through the route

$RERI_n \rightarrow RERI_{n-1} \rightarrow RERI_{n-3}$ (we name point [1]),

$\sum_{\binom{n}{n-2}}\sum_{\binom{n-2}{n-3}} RR_{(n-3)} \rightarrow (2)_{RR_{(n-3)}}$, i.e. the part of the summation that $RR_{(n-3)}$ is calculated from $RERI_{n-3}$through the route

$RERI_n \rightarrow RERI_{n-2} \rightarrow RERI_{n-3}$ (we name point [2]),

$\sum_{\binom{n}{n-3}} RR_{(n-3)} \rightarrow (\emptyset)_{RR_{(n-3)}}$, i.e. the part of the summation that $RR_{(n-3)}$ is calculated from $RERI_{n-3}$ through the direct route

$RERI_n \rightarrow RERI_{n-3}$ (we name no point [Ø]),

We also name the summation $\{1\ \ 2\}_{RR_{(n-3)}}$ as follows

$$\{1\ \ 2\}_{RR_{(n-3)}} = (1\ \ 2)_{RR_{(n-3)}} - (1)_{RR_{(n-3)}} - (2)_{RR_{(n-3)}} + (\emptyset)_{RR_{(n-3)}} = \sum_{\binom{n}{n-1}} \sum_{\binom{n-1}{n-2}} \sum_{\binom{n-2}{n-3}} RR_{(n-3)} - \sum_{\binom{n}{n-1}} \sum_{\binom{n-1}{n-3}} RR_{(n-3)} - \sum_{\binom{n}{n-2}} \sum_{\binom{n-2}{n-3}} RR_{(n-3)} + \sum_{\binom{n}{n-3}} RR_{(n-3)}$$

On the same way we have that $\{1\}_{RR_{(n-2)}} = (1)_{RR_{(n-2)}} - (\emptyset)_{RR_{(n-2)}}$, because $\{1\}_{RR_{(n-2)}} = \sum_{\binom{n}{n-1}} \sum_{\binom{n-1}{n-2}} RR_{(n-2)} - \sum_{\binom{n}{n-2}} RR_{(n-2)}$

All routes of $\{1\}_{RR_{(n-2)}}$ are multiples of $\binom{n}{n-2}$, i.e. because of (A.SC.7) we have

$(1)_{RR_{(n-2)}} = \text{sum of} \binom{n}{n-2}\ \textit{different}\ RR_{(n-2)}$, multiplied by $b_1$, where $b_1 = 2$

$(\emptyset)_{RR_{(n-2)}} = \text{sum of} \binom{n}{n-2}\ \textit{different}\ RR_{(n-2)}$, multiplied by $b_2$, where $b_2 = 1$

And on the same way, for $\{1\ \ 2\}_{RR_{(n-3)}}$, all routes of $\{1\ \ 2\}_{RR_{(n-3)}}$ are multiples of $\binom{n}{n-3}$

$(1\ \ 2)_{RR_{(n-3)}} = \text{sum of} \binom{n}{n-3}\ \textit{different}\ RR_{(n-3)}$, multiplied by $c_1$, where $c_1 = 6$

$(1)_{RR_{(n-3)}} = \text{sum of} \binom{n}{n-3}\ \textit{different}\ RR_{(n-3)}$, multiplied by $c_2$, where $c_2 = 3$

$(2)_{RR_{(n-3)}} = \text{sum of} \binom{n}{n-3}\ different\ RR_{(n-3)}$, multiplied by $c_3$, where $c_3 = 3$

$(\emptyset)_{RR_{(n-3)}} = \text{sum of} \binom{n}{n-3}\ different\ RR_{(n-3)}$, multiplied by $c_4$, where $c_4 = 1$

We further observe that the summation $\{1 \quad 2\}_{RR_{(n-3)}}$ has q=2 points and $2^q=2^2=4$ routes and are constructed as follows

a) half of the $\{1 \quad 2\}_{RR_{(n-3)}}$ routes ($=2^{2-1}$ =2 routes) have point [2] as last point. This means that

half of the $\{1 \quad 2\}_{RR_{(n-3)}}$ routes ($=2^{2-1}$ =2 routes) are made by the $\{1\}_{RR_{(n-2)}}$ route after adding point [2] by summing up all $\sum_{\binom{n-2}{n-3}} RR_{(n-3)}$, i.e.

$$(1 \quad 2)_{RR_{(n-3)}}, \text{i.e.} RERI_n \rightarrow RERI_{n-1} \rightarrow RERI_{n-2} \rightarrow RERI_{n-3} \Rightarrow \sum_{\binom{n}{n-1}} \sum_{\binom{n-1}{n-2}} \sum_{\binom{n-2}{n-3}} RR_{(n-3)} \qquad \text{(A.SC.10)}$$

and $\quad (2)_{RR_{(n-3)}} \quad \text{i.e.} = RERI_n \rightarrow RERI_{n-2} \rightarrow RERI_{n-3} \Rightarrow \sum_{\binom{n}{n-2}} \sum_{\binom{n-2}{n-3}} RR_{(n-3)} \qquad$ (A.SC.11)

We name $\{1 \quad 2\}_{RR_{(n-3);[i]}}$ the routes of $\{1 \quad 2\}_{RR_{(n-3)}}$ having as last point [i], where i=0,1 or 2.

In other words, we name $\{1 \quad 2\}_{RR_{(n-3);[i]}}$ all the routes $RERI_n \rightarrow \cdots \rightarrow RERI_{n-i} \rightarrow RERI_{n-3}$

For i=2 we find that

since $\{1\}_{RR_{(n-2)}} = \sum_{\binom{n}{n-2}} RR_{(n-2)}$, and we have that $\{1 \quad 2\}_{RR_{(n-3);[2]}} = \sum_{\binom{n}{n-2}} \sum_{\binom{n-2}{n-3}} RR_{(n-3)} = \binom{3}{1} * \sum_{\binom{n}{n-3}} RR_{(n-3)}$

$$\{1\ \ 2\}_{RR_{(n-3);[2]}} = \sum_{\binom{n}{n-1}} \sum_{\binom{n-1}{n-2}} \sum_{\binom{n-2}{n-3}} RR_{(n-3)} - \sum_{\binom{n}{n-2}} \sum_{\binom{n-2}{n-3}} RR_{(n-3)} = \sum_{\binom{n}{n-1}} \sum_{\binom{n-1}{n-2}} \sum_{\binom{n-2}{n-3}} RR_{(n-3)} - \sum_{\binom{n}{n-2}} \sum_{\binom{n-2}{n-3}} RR_{(n-3)} =$$

But due to (A.SC.7), we have that

$$= 2 * \sum_{\binom{n}{n-2}} \sum_{\binom{n-2}{n-3}} RR_{(n-3)} - \sum_{\binom{n}{n-2}} \sum_{\binom{n-2}{n-3}} RR_{(n-3)} = \sum_{\binom{n}{n-2}} \sum_{\binom{n-2}{n-3}} RR_{(n-3)}$$

b) $2^{2-2}$ =1 route of $\{1\ \ 2\}_{RR_{(n-3)}}$ is constructed having as last point [1], i.e. $\{1\ \ 2\}_{RR_{(n-3);[1]}}$

$$\{1\ \ 2\}_{RR_{(n-3);[1]}} = (1)_{RR_{(n-3)}}, \text{i.e.}\, RERI_n \rightarrow RERI_{n-1} \rightarrow RERI_{n-3} \rightarrow \sum_{\binom{n}{n-1}} \sum_{\binom{n-1}{n-3}} RR_{(n-3)} \qquad \text{(A.SC.12)}$$

Since $\{\emptyset\}_{RR_{(n-1)}} = \sum_{\binom{n}{n-1}} RR_{(n-1)}$, we have that $\{1\ \ 2\}_{RR_{(n-3);[1]}} = \sum_{\binom{n}{n-1}} \sum_{\binom{n-1}{n-3}} RR_{(n-3)} = \binom{3}{1} * \sum_{\binom{n}{n-3}} RR_{(n-3)}$

c) 1 more route of $\{1\ \ 2\}_{RR_{(n-3)}}$ is constructed having as last point [0], i.e. $\{1\ \ 2\}_{RR_{(n-3);[1]}}$

$$\{1\ \ 2\}_{RR_{(n-3);[0]}} = (\emptyset)_{RR_{(n-3)}}\, \text{i.e.}\, RERI_n \rightarrow RERI_{n-1} \rightarrow RERI_{n-3} \rightarrow \sum_{\binom{n}{n-3}} RR_{(n-3)}, \qquad \text{(A.SC.13)}$$

So we have that

$$\{1\ \ 2\}_{RR_{(n-3)}} = \sum_{j=1}^{3} (-1)^{(j-1)} \sum_{\binom{n}{n-(3-j)}} \sum_{\binom{n-(3-j)}{n-3}} RR_{(n-3)} \qquad \text{(A.SC. 14)}$$

Now, we are ready to solve (A.SC.3), by applying complete induction to show that the summation of all $\mathrm{RR}_{(n-k)}$ , for all k=1,2,...n when we replace all $\mathrm{RERI}_{(n)}$ up to $\mathrm{RERI}_{(n-k)}$ in (A.SC.3) is $\sum_{\binom{n}{n-k}} \mathrm{RR}_{(n-k)}$

In order to prove that, we apply step by step complete induction

1) for i=1, the summation for $\mathrm{RR}_{(n-1)}$ in (A.SC.3) is (see (A.SC.4))

$$\sum_{\binom{n}{n-1}} \mathrm{RR}_{(n-1)}$$

i.e. our hypothesis hold

2) for all i=2,3,...k-1, the summation for $\mathrm{RR}_{(n-i)}$ in (A.SC.3) is

$$\sum_{\binom{n}{n-i}} \mathrm{RR}_{(n-i)}$$

3) for i=k, we have to show that the summation for $\mathrm{RR}_{(n-k)}$ when we replace all $\mathrm{RERI}_{(n)}$ up to $\mathrm{RERI}_{(n-k)}$ in (A.SC.3) is

$$\sum_{\binom{n}{n-k}} \mathrm{RR}_{(n-k)} \qquad \text{(A.SC.15)}$$

To prove (A.SC.15), we have to show that

$$\{1 \quad 2 \quad 3 \quad \dots \quad (k-2) \quad (k-1)\}_{\mathrm{RR}_{(n-k);[i]}} = \sum_{\binom{n}{n-i}} \sum_{\binom{n-i}{n-k}} \mathrm{RR}_{(n-k)} = \binom{k}{i} * \sum_{\binom{n}{n-k}} \mathrm{RR}_{(n-k)} \qquad (\mathrm{A.SC.16})$$

In other words, from (A.SC.16), we have all the different $\mathrm{RR}_{(\mathrm{n-k})}$ appear $\binom{\mathrm{k}}{\mathrm{i}}$ times from the routes of

$\{1 \;\; 2 \;\; 3 \;\; \ldots \;\; (k-2) \;\; (k-1)\}_{\mathrm{RR}_{(\mathrm{n-k})}}$ having as last point [i], i.e. through all the routes $\mathrm{RERI_n} \to \cdots \to \mathrm{RERI_{n-i}} \to \mathrm{RERI_{n-k}}$

To prove (A.SC.16), we have to show that from the total $\sum_{\binom{\mathrm{n}}{\mathrm{n-i}}} \mathrm{RR}_{(\mathrm{n-i})}$ of all different $\mathrm{RR}_{(\mathrm{n-i})}$, we end up to $\binom{\mathrm{k}}{\mathrm{i}} * \sum_{\binom{\mathrm{n}}{\mathrm{n-k}}} \mathrm{RR}_{(\mathrm{n-k})}$ different $\mathrm{RR}_{(\mathrm{n-k})}$.

We have that

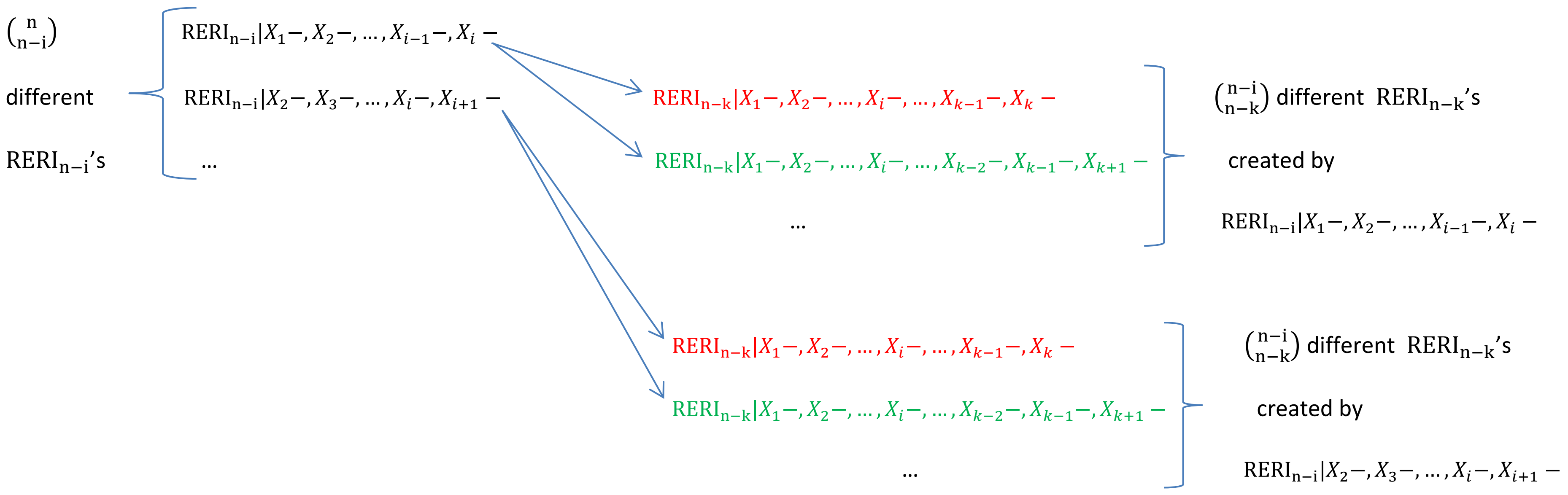

From above, we can observe that all the $\binom{n}{n-k}$ different $\mathrm{RERI}_{n-k}$ are created from all the $\binom{n}{n-i}$ different $\mathrm{RERI}_{n-i}$'s $\binom{k}{i}$ times. This is done because, for p∈P={1,.2,...,k}, $\mathrm{RERI}_{n-k}|X_1 -, X_2 -, \dots, X_p -, \dots, X_k -$ is derived by all $\mathrm{RERI}_{n-i}|i\ of\ the\ X_k = 0$, with i≤k, or in other words, i∈I⊂P, but from no other $\mathrm{RERI}_{n-i}|any\ X_q -$, for q∉ P={1,.2,...,k}. The number of sets of I's , for which I⊂P is $\binom{k}{i}$. So the total $\mathrm{RR}_{(n-k)}$ that will be created through the route $\mathrm{RERI}_n \to \cdots \to \mathrm{RERI}_{n-i} \to \cdots \to \mathrm{RERI}_{n-k}$ is equal to $\sum_{\binom{n}{n-i}} \sum_{\binom{n-i}{n-k}} \mathrm{RR}_{(n-k)} = \binom{k}{i} * \sum_{\binom{n}{n-k}} \mathrm{RR}_{(n-k)}$ . So (A.SC.16) holds.

Now, using (A.SC.16), we have that

$\{1 \quad 2 \quad 3 \quad \dots \quad (k-2) \quad (k-1)\}_{\mathrm{RR}_{(n-k)}}$ is constructed by

$$\{1 \quad 2 \quad 3 \quad \dots \quad (k-2) \quad (k-1)\}_{\mathrm{RR}_{(n-k);[k-1]}} = \sum_{\binom{n}{n-(k-1)}} \sum_{\binom{n-(k-1)}{n-k}} \mathrm{RR}_{(n-k)} = \sum_{\binom{n}{n-k+1}} \sum_{\binom{n-k+1}{n-k}} \mathrm{RR}_{(n-k)} = \binom{k}{k-1} \sum_{\binom{n}{n-k}} \mathrm{RR}_{(n-k)}$$

$$\{1 \quad 2 \quad 3 \quad \dots \quad (k-2) \quad (k-1)\}_{\mathrm{RR}_{(n-k);[k-2]}} = \sum_{\binom{n}{n-(k-2)}} \sum_{\binom{n-(k-2)}{n-k}} \mathrm{RR}_{(n-k)} = \sum_{\binom{n}{n-k+2}} \sum_{\binom{n-k+2}{n-k}} \mathrm{RR}_{(n-k)} = \binom{k}{k-2} \sum_{\binom{n}{n-k}} \mathrm{RR}_{(n-k)}$$

...

$$\{1\quad 2\quad 3\quad \dots\quad (k-2)\quad (k-1)\}_{\mathrm{RR}_{(n-k);[1]}} = \sum_{\binom{n}{n-(k-(k-1))}} \sum_{\binom{n-(k-(k-1))}{n-k}} \mathrm{RR}_{(n-k)} = \sum_{\binom{n}{n-1}} \sum_{\binom{n-1}{n-k}} \mathrm{RR}_{(n-k)} = \binom{k}{1} \sum_{\binom{n}{n-k}} \mathrm{RR}_{(n-k)}$$

$$\{1\quad 2\quad 3\quad \dots\quad (k-2)\quad (k-1)\}_{\mathrm{RR}_{(n-k);[\emptyset]}} = \sum_{\binom{n}{n-(k-(k))}} \sum_{\binom{n-(k-(k))}{n-k}} \mathrm{RR}_{(n-k)} = \sum_{\binom{n}{n}} \sum_{\binom{n}{n-k}} \mathrm{RR}_{(n-k)} = \binom{k}{0} \sum_{\binom{n}{n-k}} \mathrm{RR}_{(n-k)}$$

and as we can see from (A.SC.13), in a way that

$$\begin{aligned}\{1\quad 2\quad 3\quad \dots\quad (k-2)\quad (k-1)\}_{\mathrm{RR}_{(n-k)}} = {} & \{1\quad 2\quad 3\quad \dots\quad (k-2)\quad (k-1)\}_{\mathrm{RR}_{(n-k);[k-1]}} \\ & -\{1\quad 2\quad 3\quad \dots\quad (k-2)\quad (k-1)\}_{\mathrm{RR}_{(n-k);[k-2]}} \\ & +\{1\quad 2\quad 3\quad \dots\quad (k-2)\quad (k-1)\}_{\mathrm{RR}_{(n-k);[k-3]}} \\ & \dots \\ & \pm\{1\quad 2\quad 3\quad \dots\quad (k-2)\quad (k-1)\}_{\mathrm{RR}_{(n-k);[\emptyset]}} \qquad \text{(A.SC.17)}\end{aligned}$$

So extending (A.SC.14), we have that

$$\{1\quad 2\quad 3\quad \dots\quad (k-2)\quad (k-1)\}_{\mathrm{RR}_{(n-k)}} = \sum_{j=1}^{k} (-1)^{(j-1)} \sum_{\binom{n}{n-(k-j)}} \sum_{\binom{n-(k-j)}{n-k}} \mathrm{RR}_{(n-k)}$$

So, as exactly we worked to prove (A.SC.7), we have that

$$\{1 \quad 2 \quad 3 \quad \ldots \quad (k-2) \quad (k-1)\}_{\mathrm{RR}_{(\mathrm{n-k})}} = \sum_{j=1}^{k} (-1)^{j-1} \binom{\mathrm{k}}{\mathrm{k}-(\mathrm{k}-\mathrm{j})} \sum_{\binom{\mathrm{n}}{\mathrm{n-k}}} \mathrm{RR}_{(\mathrm{n-k})} = \sum_{\binom{\mathrm{n}}{\mathrm{n-k}}} \mathrm{RR}_{(\mathrm{n-k})} * \sum_{j=1}^{k} (-1)^{j-1} \binom{\mathrm{k}}{\mathrm{j}}$$

$$= \sum_{\binom{\mathrm{n}}{\mathrm{n-k}}} \mathrm{RR}_{(\mathrm{n-k})} * \left( \binom{\mathrm{k}}{0} - \sum_{j=0}^{k} (-1)^{j} \binom{\mathrm{k}}{\mathrm{j}} \right)$$

$$= \sum_{\binom{\mathrm{n}}{\mathrm{n-k}}} \mathrm{RR}_{(\mathrm{n-k})} * \left( \binom{\mathrm{k}}{0} - \sum_{j=0}^{k} \binom{\mathrm{k}}{\mathrm{j}} (1)^{k-j} * (-1)^{j} \right)$$

$$= \sum\nolimits_{\binom{\mathrm{n}}{\mathrm{n-k}}} \mathrm{RR}_{(\mathrm{n-k})} * (1 - (1-1)^{k}) = \sum\nolimits_{\binom{\mathrm{n}}{\mathrm{n-k}}} \mathrm{RR}_{(\mathrm{n-k})}$$

The transformation in the last line was applied because

$\sum_{j=0}^{k} \binom{\mathrm{k}}{\mathrm{j}} (x)^{k-j} * (y)^{j} = (x+y)^{k}$, for x=1, y=-1

Of note that the same hold for k=n-1 and k=n for $\mathrm{RR}_{(1)}$, $\mathrm{RR}_{(0)}$ respectively i.e. the total summation of $\mathrm{RR}_{(1)}$ and $\mathrm{RR}_{(0)}$ after replacing all $\mathrm{RERI}_{(\mathrm{n})}$ up to $\mathrm{RERI}_{(\mathrm{n-2})}$ from the recurrence relation (A.SC.3) and $\mathrm{ERR}_{(1)} = \mathrm{RR}_{(1)} - \mathrm{RR}_{(0)}$,

$$\sum_{j=1}^{n-1}(-1)^{j-1}\binom{\mathrm{n}-1}{\mathrm{n}-1-(\mathrm{n}-1-\mathrm{j})}\sum_{\binom{\mathrm{n}}{\mathrm{n}-(\mathrm{n}-1)}}\mathrm{RR}_{(\mathrm{n}-(\mathrm{n}-1))}=\sum_{\binom{\mathrm{n}}{1}}\mathrm{RR}_{(1)}$$

and

$$\sum_{j=1}^{n}(-1)^{j-1}\binom{\mathrm{n}}{\mathrm{n}-(\mathrm{n}-\mathrm{j})}\sum_{\binom{\mathrm{n}}{\mathrm{n}-(\mathrm{n})}}\mathrm{RR}_{(\mathrm{n}-(\mathrm{n}))}=\mathrm{RR}_{(0)}$$

We also note from the (A.SC.17) that in the final solution, the sign changes from + to - (or vice versa) before each

$$\sum_{\binom{\mathrm{n}}{\mathrm{k}}}\mathrm{RR}_{(\mathrm{k})}$$

So we have that the final solution for the recurrence relationship (A.SC.3) is

$$\mathrm{RERI}_{\mathrm{n}}(\mathrm{X}_1,\mathrm{X}_2,\dots,\mathrm{X}_{\mathrm{n}})=\mathrm{RR}_{(\mathrm{n})}$$

$$-\sum_{\binom{\mathrm{n}}{\mathrm{n}-1}}\mathrm{RR}_{(\mathrm{n}-1)}$$

$$+\sum_{\binom{\mathrm{n}}{\mathrm{n}-2}}\mathrm{RR}_{(\mathrm{n}-2)}$$

...

$$+(-1)^{n} * \sum_{\binom{n}{0}} \mathrm{RR}_{(0)}$$

or in other words

$$\mathrm{RERI}_{n}(X_1, X_2, \dots, X_n) = \sum_{k=0}^{n} \sum_{\binom{n}{n-k}} (-1)^{k}\, \mathrm{RR}_{(n-k)} \qquad \textbf{\textit{QED}}\ (\textbf{\textit{see equation 6 in the text}})$$

*Proof of expression (6)*

Connection between n-way interaction and (n-1) way interactions

We have to prove expression (7) from the paper, that is

$$\mathrm{RERI_n}(X_1, X_2, \dots, X_n) = \left(\mathrm{RERI}_{n-1}(X_1, X_2, \dots, X_n | X_i = 1) * \ \mathrm{RR}_{X_1 - X_2 - \dots X_{i+1} - X_i + X_{i+1} - \cdots X_{i+1} - \cdots X_n -}\right) - \mathrm{RERI}_{n-1}(X_1, X_2, \dots, X_n | X_i = 0)$$

From equation 6 in the paper, we have that

$$\mathrm{RERI_n}(X_1, X_2, \dots, X_n) = \mathrm{RR_{(n)}} - \sum_{\binom{n}{n-1}} \mathrm{RR_{(n-1)}} + \sum_{\binom{n}{n-2}} \mathrm{RR_{(n-2)}} - \sum_{\binom{n}{n-3}} \mathrm{RR_{(n-3)}} + \cdots + (-1)^n * \sum_{\binom{n}{0}} \mathrm{RR_{(0)}}$$

Using expression (6), we can calculate

$$\mathrm{RERI}_{n-1}(X_1, X_2, \dots, X_n | X_i = 1) = \frac{\mathrm{RR}_{(n);X_i+} - \sum_{\binom{n-1}{1}} \mathrm{RR}_{(n-1);X_i+} + \sum_{\binom{n-1}{2}} \mathrm{RR}_{(n-2);X_i+} \dots + (-1)^{n-1} * \sum_{\binom{n-1}{n-1}} \mathrm{RR}_{(1);X_i+}}{\mathrm{RR}_{X_1 - X_2 - \dots X_{i+1} - X_i + X_{i+1} - \cdots X_{i+1} - \cdots X_n}} \quad \text{(A.SC.18)}$$

where $\mathrm{RR}_{(k);X_i+}$ , k≤n and i≤n, is the relative risk, when k out of n risk factors are present and among them $X_i$ is present as well.

In (A.SC.18), using the notation $\mathrm{RR}_{(k);X_i+}$ , we denote the RR when k risk factors are present and n-k are absent and $X_i$ is present

From the above, it is obvious that the term $\mathrm{RR}_{(1);X_i+}$ is equal to $\mathrm{RR}_{(1);X_i+} = \mathrm{RR}_{X_1 - X_2 - \dots X_{i+1} - X_i + X_{i+1} - \cdots X_{i+1} - \cdots X_n -}$

So, from (A.SC.18), we have that

$$\mathrm{RERI}_{n-1}(X_1, X_2, \dots, X_n | X_i = 1) * \mathrm{RR}_{X_1 - X_2 - \dots X_{i+1} - X_i + X_{i+1} - \cdots X_{i+1} - \cdots X_n -} =$$

$$= RR_{(n);X_i+} - \sum_{\binom{n-1}{1}} RR_{(n-1);X_i+} + \sum_{\binom{n-1}{2}} RR_{(n-2);X_i+} \dots + (-1)^{n-1} * \sum_{\binom{n-1}{n-1}} RR_{(1);X_i+} \qquad \text{(A.SC.19)}$$

In the same fashion, due to expression 6 we have that

$$RERI_{n-1}(X_1, X_2, \dots, X_n | X_i = 0) = RR_{(n-1);X_i-} - \sum_{\binom{n-1}{n-2}} RR_{(n-2);X_i-} + \sum_{\binom{n-1}{n-3}} RR_{(n-3);X_i-} \dots + (-1)^{n-1} * \sum_{\binom{n-1}{0}} RR_{(0);X_i-} \qquad \text{(A.SC.20)}$$

where $RR_{(k);X_i-}$ , k≤n and i≤n, is the relative risk, when k out of n risk factors are present and risk factor $X_i$ is absent.

Now, combining (A.SC.19) and (A.SC.20), we have that

$$\left(RERI_{n-1}(X_1, X_2, \dots, X_n | X_i = 1) * RR_{X_1 - X_2 - \dots X_{i+1} - X_i + X_{i+1} - \cdots X_{i+1} - \cdots X_n -}\right) - RERI_{n-1}(X_1, X_2, \dots, X_n | X_i = 0) =$$

$$= RR_{(n);X_i+} - \sum_{\binom{n-1}{1}} RR_{(n-1);X_i+} + \sum_{\binom{n-1}{2}} RR_{(n-2);X_i+} \dots + (-1)^{n-1} * \sum_{\binom{n-1}{n-1}} RR_{(1);X_i+}$$

$$-\left(RR_{(n-1);X_i-} - \sum_{\binom{n-1}{n-2}} RR_{(n-2);X_i-} + \sum_{\binom{n-1}{n-3}} RR_{(n-3);X_i-} \dots + (-1)^{n-1} * \sum_{\binom{n-1}{0}} RR_{(0);X_i-}\right)$$

Now if we try to summarize all $RR_{(n-j)}$ terms (1≤j≤n). We have that

$$\left(RERI_{n-1}(X_1, X_2, \dots, X_n | X_i = 1) * RR_{X_1 - X_2 - \dots X_{i+1} - X_i + X_{i+1} - \cdots X_{i+1} - \cdots X_n -}\right) - RERI_{n-1}(X_1, X_2, \dots, X_n | X_i = 0) = RR_{(n);X_i+}$$

$$-\left(\sum_{\binom{n-1}{n-2}} RR_{(n-1);X_i+} + RR_{(n-1);X_i-}\right)$$

$$+\left(\sum_{\binom{n-1}{n-3}} RR_{(n-1);X_i+} + \sum_{\binom{n-1}{n-2}} RR_{(n-1);X_i-}\right)$$

$$-\left(\sum_{\binom{n-1}{n-4}} RR_{(n-2);X_i+} + \sum_{\binom{n-1}{n-3}} RR_{(n-2);X_i-}\right)$$

$$\dots$$

$$+(-1)^{j}\left(\sum_{\binom{n-1}{n-j}} RR_{(n-j);X_i+} + \sum_{\binom{n-1}{n-(j-1)}} RR_{(n-j);X_i-}\right)$$

$$\ldots$$

$$+(-1)^{n-2}\left(\sum_{\binom{n-1}{1}} RR_{(2);X_i+} + \sum_{\binom{n-1}{2}} RR_{(2);X_i-}\right)$$

$$+(-1)^{n-1}\left(\sum_{\binom{n-1}{0}} RR_{(1);X_i+} + \sum_{\binom{n-1}{1}} RR_{(1);X_i-}\right)$$

$$+(-1)^{n}\left(\sum_{\binom{n-1}{0}} RR_{(0);X_i-}\right) \qquad \text{(A.SC.21)}$$

It is obvious from (A.SC.21), that if we prove for all $RR_{(n-j)}$ terms ($1 \leq j \leq n$) that

$$+(-1)^{j}\left(\sum_{\binom{n-1}{n-j}} RR_{(n-j);X_i+} + \sum_{\binom{n-1}{n-(j-1)}} RR_{(n-j);X_i-}\right) = +(-1)^{j}\left(\sum_{\binom{n}{n-j}} RR_{(n-j)}\right) \qquad \text{(A.SC.22)}$$

then we will have also proved equation (7) in the paper.

With a closer look at the first part of (A.SC.22), we have that $\sum_{\binom{n-1}{n-j}} RR_{(n-j);X_i+}$ and $\sum_{\binom{n-1}{n-(j-1)}} RR_{(n-j);X_i-}$ share no common RRs, given that $\sum_{\binom{n-1}{n-j}} RR_{(n-j);X_i+}$ is the summation of all the combinations of different RRs, when n-j risk factors are present and j are absent <u>and $X_i$ is present</u>, while $\sum_{\binom{n-1}{n-(j-1)}} RR_{(n-j);X_i-}$ is the summation of all the combinations of different RRs, when n-j risk factors are present and j are absent <u>and $X_i$ is absent</u>, so there is no overlapping of common RRs in that summation. Moreover, it is know from the Pascal rule that

$$\binom{n-1}{n-j} + \binom{n-1}{n-(j-1)} = \binom{n}{n-j}$$

So the summation of $\sum_{\binom{n-1}{n-j}} RR_{(n-j);X_i+}$ and $\sum_{\binom{n-1}{n-(j-1)}} RR_{(n-j);X_i-}$ will be equal to $\sum_{\binom{n}{n-j}} RR_{(n-j)}$, i.e. the summation of all the combinations of different RRs, when n-j risk factors are present and j are absent. ***QED***

So (A.SC.22) is proven, which means that expression (7) in the paper is proven as well.

## Section D - Clarifications and recommendations for calculating multi-way interactions

When dealing only with two risk factors, the definition of RERI is pretty straightforward. The only issue that we should take into consideration is that RERI is calculated by risk and not protective factors. So if any of the binary variables $Z_1$ or $Z_2$ we use in the analysis is a protective factor, we need to recode them to a risk factor. For example, if $Z_1$ and $Z_2$ are both protective factors, we need to create variables $X_1=1-Z_1$ and $X_2=1-Z_2$ and calculate RERI($X_1$,$X_2$). Nevertheless, when it is of interest the calculation of higher order interactions on the additive scale, we showed that the definitions are getting more complex, so we will try to further elaborate on them.

### Three-way interactions

To shed more light in the properties of joint effects of the 3 risk factors of interest, we propose the following steps;

1) Conduct the analysis with the exposure of interest $Z_1$, $Z_2$, $Z_3$, without using any interaction term. We run e.g. Cox regression, and we check if the hazard ratios (HR) are greater than one (i.e. if the exposures are risk factors). If, for example, we have that HR($Z_1$)>1, HR($Z_2$)>1, but HR($Z_3$)<1, (i.e. $Z_1$, $Z_2$ are risk factors, but $Z_3$ is protective), we have to use the variables $X_1=Z_1$, $X_2=Z_2$, and $X_3=1-Z_3$. In our illustrative paradigm, we didn't use a variable for the adherence to MD, which is a protective factor for mortality, bur for non-adherence to MD instead.

2) Perform the analysis with the risk factors of interest $X_1$, $X_2$, $X_3$, this time using the interaction terms $X_1X_2$, $X_1X_3$, $X_2X_3$ and $X_1X_2X_3$, as presented in model (A.SB.1) in the Appendix (Section B)
3) Compute $TotRERI_3(X_1,X_2,X_3)$ (from A.SB.2, Appendix, Section B) and find out whether the effects of $X_1$, $X_2$, $X_3$ are super-additive ($TotRERI_3(X_1,X_2,X_3)>0$), additive ($TotRERI_3(X_1,X_2,X_3)=0$) or sub-additive ($TotRERI_3(X_1,X_2,X_3)<0$). In our example, we show that the effects of low MD, obesity, and current smoking are super-additive ($TotRERI_3(X_1,X_2,X_3)=1.20>0$).
4) Estimate $RERI_3(X_1,X_2,X_3)$ (from A.SB.3, Appendix, Section B) to check whether any deviation from additivity of the three risk factors ($TotRERI_3(X_1,X_2,X_3)$) is attributed to the 3-way interaction, beyond the two way interactions, given that the 3$^{rd}$ risk factor is absent [see equation (3)]. In our example, there was evidence for 3-way additive interaction of the risk factors, beyond their two way interactions.
5) Calculate $RERI_2(X_1,X_2|X_3=0)$, $RERI_2(X_1,X_3|X_2=0)$ and $RERI_2(X_2,X_3|X_1=0)$ [from (A.SB.4) - (A.SB.6), Appendix, Section B) to test whether any deviation from additivity of the three risk factors (expressed through $TotRERI_3(X_1,X_2,X_3)$) is attributed to additive interaction of the two risk factors [(see equation (3)], when the third is absent. In our paradigm, all $RERI_2$ given the absence of the 3$^{rd}$ risk factor were negative, but of small magnitude and not statistically significant.
6) Compute $RERI_2(X_1,X_2|X_3=1)$, $RERI_2(X_1,X_3|X_2=1)$ and $RERI_2(X_2,X_3|X_1=1)$ (from (A.SB.7) – (A.SB.9) in Appendix, Section B) to check to what extend the interaction due to three risk factors exclusively (i.e $RERI_3(X_1,X_2,X_3)$) can be interpreted as interaction of the two risk factors, given that the 3$^{rd}$ is present. If we combine that information with the one from step 5, we can additionally check whether $RERI_2$'s remain constant across the strata of the 3rd risk factor of interest. In our paradigm, we observed that all $RERI_2$ were positive, given the presence of the 3rd risk factor. If we additionally take into consideration $RERI_2$'s given the absence of the 3rd risk factor (from the previous step), we draw the conclusion that there is difference in how two of these variables interact across the levels of the 3rd factor.

7) Check for qualitative interaction, i.e. whether

a) the risk of $X_1$ is increasing across the strata of $X_2$ and $X_3$, i.e.

$RR_{X_1+X_2+X_3+} > RR_{X_1-X_2+X_3+}$ , $RR_{X_1+X_2+X_3-} > RR_{X_1-X_2+X_3-}$,

$RR_{X_1+X_2-X_3+} > RR_{X_1-X_2-X_3+}$ and $RR_{X_1+X_2-X_3-} > RR_{X_1-X_2-X_3-}$

b) the risk of $X_2$ is increasing across the strata of $X_1$ and $X_3$, i.e.

$RR_{X_1+X_2+X_3+} > RR_{X_1+X_2-X_3+}$ , $RR_{X_1+X_2+X_3-} > RR_{X_1+X_2-X_3-}$ ,

$RR_{X_1-X_2+X_3+} > RR_{X_1-X_2-X_3+}$ and $RR_{X_1-X_2+X_3-} > RR_{X_1-X_2-X_3-}$

c) the risk of $X_3$ is increasing across strata of $X_1$ and $X_2$, i.e.

$RR_{X_1+X_2+X_3+} > RR_{X_1+X_2+X_3-}$ , $RR_{X_1+X_2-X_3+} > RR_{X_1+X_2-X_3-}$ ,

$RR_{X_1-X_2+X_3+} > RR_{X_1-X_2+X_3-}$ and $RR_{X_1-X_2-X_3+} > RR_{X_1-X_2-X_3-}$

For more details on qualitative interaction, please see Section F

n-way interactions

The total relative excess risk due to interaction ($TotRERI_n$) is calculated by comparing the joint effect of all n risk factors to the situation when each one acts separately (expression 5). However, $TotRERI_n$ is affected by all lower-order interactions of the n risk factors and not exclusively by the n-way interaction of the risk factors. From equation (A.SC.2), we observe that the sign and magnitude of $TotRERI_n$ depends on the sign and magnitude of $\sum_{\binom{n}{n-1}} RERI_{n-1}(X_1, X_2, \dots, X_n | 1 \text{ of the } X_i = 0)$, $\sum_{\binom{n}{n-2}} RERI_{n-2}(X_1, X_2, \dots, X_n | 2 \text{ of the } X_i = 0)$, …, $\sum_{\binom{n}{2}} RERI_2(X_1, X_2, \dots, X_n | (n-2) \text{ of the } X_i = 0)$. So, $RERI_n(X_1, X_2, \dots, X_n)$ measures the interaction between n risk factors on the additive scale, as this index does not account for all the lower order additive interactions. To shed more light in the properties of joint effects of the n risk factors, we propose the following steps;

1) Apply step 1 as in recommendations in the 3-way interactions section, this time for $Z_1$, $Z_2$, …, $Z_n$. From this step, we will end up with the risk factors $X_1$, $X_2$, …, $X_n$.
2) Include all possible 2,3,…,n product terms constructed by $X_1$, $X_2$, …, $X_n$ in the model, as described in step 2 in recommendations in the 3way interactions section.
3) Calculate $TotRERI_n(X_1, X_2, \dots, X_n)$ from expression (5).
4) Calculate $RERI_n(X_1, X_2, \dots, X_n)$ from equation (6)
5) Compute all $RERI_{n-1}$ given the 1 risk factor is absent ($RERI_{n-1}(X_1, X_2, \dots, X_n | 1 \text{ of the } X_i = 0)$) from expression (6).
6) Estimate all $RERI_{n-1}$ given the 1 risk factor is present ($RERI_{n-1}(X_1, X_2, \dots, X_n | 1 \text{ of the } X_i = 1)$) from equation (A.SC.19)
7) We can additionally compute all $TotRERI_{n-k}$ and $RERI_{n-k}$, $2 \le k \le n-2$, given k risk factors are all absent ($TotRERI_{n-k}(X_1, X_2, \dots, X_n | k \text{ of the } X_i = 0)$ and $RERI_{n-k}(X_1, X_2, \dots, X_n | k \text{ of the } X_i = 0)$) to understand how and under which conditions the n-k risk factors interact for the development of a specific disease. On the same fashion, we can compute all $TotRERI_{n-k}(X_1, X_2, \dots, X_n | k \text{ of the } X_i = 1)$ and $RERI_{n-k}(X_1, X_2, \dots, X_n | k \text{ of the } X_i = 1)$

8) Check for qualitative interaction, that is

whether the risk of $X_1$ is increasing across the strata of $X_2$, $X_3$ … $X_n$,.

whether the risk of $X_2$ is increasing across the strata of $X_1$, $X_3$ … $X_n$,

…

whether the risk of $X_n$ is increasing across the strata of $X_1$, $X_2$ … $X_{n-1}$

## Section E – Multiplicative interaction and its connection to additive interaction

2-way interactions

The usual practice of the researchers is to refer to statistical interaction when studying interaction between risk factors. Nevertheless, under this concept, interaction is measured on either additive or multiplicative scale, depending on the form of the underlying model used. It is known in the study of the 2-way interactions that then we use Cox or logistic regression, which are inherently multiplicative models, then the beta coefficient of the product term shows whether there is any deviation from the multiplicativity of the effects of two risk factors. More specifically, the effects are super- or sub-multiplicative, if the beta coefficient is greater or lower than zero respectively (or, equivalently, if the odds/hazard ratio is higher or lower than 1). For example, if we run a Cox regression model with exposures $X_1$ and $X_2$, i.e.

$$\lambda(t) = \lambda_0(t) * \exp(b_1X_1 + b_2X_2 + b_3X_1X_2)\,, \quad \text{(A.SE.1)}$$

then, we calculate

$$HR(X_1 = 0, X_2 = 0) = RR_{X_1-X_2-} = 1$$

$$HR(X_1 = 1, X_2 = 0) = RR_{X_1+X_2-} = \exp(b_1)$$

$$HR(X_1 = 0, X_2 = 1) = RR_{X_1-X_2+} = \exp(b_2)$$

$$HR(X_1 = 1, X_2 = 1) = RR_{X_1+X_2+} = \exp(b_1 + b_2 + b_3)$$

Then the multiplicative interaction

$$I_2 = \frac{RR_{X_1+X_2+}}{RR_{X_1+X_2-} * RR_{X_1-X_2+}} = \exp(b_3)\,, \qquad \text{(A.SE.2)}$$

So from (A.SE.2), the effects are super-multiplicative ($I_2>1$), if $b_3>0$,

the effects are multiplicative ($I_2$=1), if $b_3$=0 and

the effects are sub-multiplicative ($I_2$<1), if $b_3$<0,

that is, the statistical interaction will show whether there is any deviation from multiplicativity of the effects.

Regarding the connection between multiplicative and additive 2-way interaction, there are 2 inequalities that link the deviation from additivity and from multiplicativity. Both of them hold when there is no qualitative interaction. They also hold if we relax the assumptions of qualitative interaction and we assume only that $RR_{X_1+X_2-} > RR_{X_1-X_2-}(= 1)$ and

$RR_{X_1-X_2+} > RR_{X_1-X_2-}(= 1)$

*1) When the effects of 2 risk factors are either multiplicative of super-multiplicative, then the effects will be super-additive.*

Proof: We have that

$$RERI_2(X_1, X_2) = RR_{X_1+X_2+} - RR_{X_1+X_2-} - RR_{X_1-X_2+} + 1 \quad , \quad \text{(A.SE.3)}$$

In the case of multiplicative or super-multiplicative effects of $X_1$ and $X_2$, i.e. when $I_2 \geq 1$, from (A.SE.2), $RR_{X_1+X_2+} \geq RR_{X_1+X_2-} * RR_{X_1-X_2+}$, so from (A.SE.3) we have

$$RERI_2(X_1, X_2) \geq RR_{X_1+X_2-} * RR_{X_1-X_2+} - RR_{X_1+X_2-} - RR_{X_1-X_2+} + 1$$

$$= RR_{X_1+X_2-} * \left(RR_{X_1-X_2+} - 1\right) - \left(RR_{X_1-X_2+} - 1\right)$$

$$= \left(RR_{X_1+X_2-} - 1\right) * \left(RR_{X_1-X_2+} - 1\right) > 0$$

because $RR_{X_1+X_2-}$ and $RR_{X_1+X_2-} > 1$, as $X_1$ and $X_2$ are risk factors. In other words, we proved that when the effects of 2 risk actors are either multiplicative of super-multiplicative, then the effects will be super-additive.

*2) When the effects of 2 risk factors are either additive or sub-additive, then the effects will be sub-multiplicative.*

Proof: If $RERI_2(X_1, X_2) \leq 0$, then $RR_{X_1+X_2+} \leq RR_{X_1+X_2-} + RR_{X_1-X_2+} - 1$ , so

$$\frac{RR_{X_1+X_2+}}{RR_{X_1+X_2-}+RR_{X_1-X_2+}-1} \leq 1 \quad \text{(A.SE.4)},$$

so we have to prove that

$RR_{X_1+X_2-} + RR_{X_1-X_2+} - 1 < RR_{X_1+X_2-} * RR_{X_1-X_2+}$, or equivalently

$RR_{X_1+X_2-} * RR_{X_1-X_2+} - RR_{X_1+X_2-} - RR_{X_1-X_2+} + 1 > 0$ or equivalently

$\left(RR_{X_1+X_2-} - 1\right) * \left(RR_{X_1-X_2+} - 1\right) > 0$, which is true

So from (A.SE.4), we have that

$$I_2 = \frac{RR_{X_1+X_2+}}{RR_{X_1+X_2-} * RR_{X_1-X_2+}} < \frac{RR_{X_1+X_2+}}{RR_{X_1+X_2-} + RR_{X_1-X_2+} - 1} \leq 1$$

So, $I_2<1$. In other words, we proved that when the effects of 2 risk factors are either additive or sub-additive, then the effects will be sub-multiplicative.

3-way interactions

In case of study of 3-way interaction on the multiplicative scale, we should extend the definitions to three risk factors $X_1$, $X_2$ and $X_3$. We will use the worked example from our paper, that is we will use the Cox regression model (A.SB.1).

For the deviation from multiplicativity of the effects of these factors, one should compare

$$RR_{X_1+X_2+X_3+} \quad \text{vs} \quad RR_{X_1+X_2-X_3-} * RR_{X_1-X_2+X_3-} * RR_{X_1-X_2-X_3+}$$

So, we can calculate $TotI_3(X_1, X_2, X_3)$ to check if there is any deviation from multiplicativity

$$\mathrm{TotI}_3(X_1, X_2, X_3) = \frac{RR_{X_1+X_2+X_3+}}{RR_{X_1+X_2-X_3-} * RR_{X_1-X_2+X_3-} * RR_{X_1-X_2-X_3+}} \quad \text{(A.SE.5)}$$

So we have that

$$\mathrm{TotI}_3(X_1, X_2, X_3) = \frac{\exp(a_1+a_2+a_3+a_4+a_5+a_6+a_7)}{\exp(a_1)*\exp(a_2)*\exp(a_3)} = \exp(a_4 + a_5 + a_6 + a_7) \quad \text{(A.SE.6)}$$

On the same fashion, the 2-way multiplicative interaction, given the third factor is absent will be

$$I_2(X_1, X_2|X_3 = 0) = \frac{\exp(a_1+a_2+a_4)}{\exp(a_1)*\exp(a_2)} = \exp(a_4) \quad \text{(A.SE.7)}$$

$$I_2(X_1, X_3|X_2 = 0) = \frac{\exp(a_1+a_3+a_5)}{\exp(a_1)*\exp(a_3)} = \exp(a_5) \quad \text{(A.SE.8)}$$

$$I_2(X_2, X_3|X_1 = 0) = \frac{\exp(a_2+a_3+a_6)}{\exp(a_1)*\exp(a_2)} = \exp(a_6) \quad \text{(A.SE.9)}$$

Moreover, the 2-way multiplicative interaction, given the third factor is present will be

$$I_2(X_1, X_2|X_3 = 1) = \frac{\exp(a_1+a_2+a_3+a_4+a_5+a_6+a_7)/\exp(a_3)}{(\frac{\exp(a_1+a_3+a_5)}{\exp(a_3)})*(\frac{\exp(a_2+a_3+a_6)}{\exp(a_3)})} = \exp(a_4 + a_7) \quad \text{(A.SE.10)}$$

$$I_2(X_1, X_3|X_2 = 1) = \frac{\exp(a_1+a_2+a_3+a_4+a_5+a_6+a_7)/\exp(a_2)}{(\frac{\exp(a_1+a_2+a_4)}{\exp(a_2)})*(\frac{\exp(a_2+a_3+a_6)}{\exp(a_2)})} = \exp(a_5 + a_7) \quad \text{(A.SE.11)}$$

$$I_2(X_2, X_3|X_1 = 1) = \frac{\exp(a_1+a_2+a_3+a_4+a_5+a_6+a_7)/\exp(a_1)}{(\frac{\exp(a_1+a_2+a_4)}{\exp(a_1)})*(\frac{\exp(a_1+a_3+a_5)}{\exp(a_1)})} = \exp(a_6 + a_7) \quad \text{(A.SE.12)}$$

and the 3-way multiplicative interaction, beyond the 2-way interactions will be

$$I_3(X_1, X_2, X_3) = \frac{\mathrm{TotI}_3(X_1, X_2, X_3)}{I_2(X_1, X_2|X_3 = 0) * I_2(X_1, X_3|X_2 = 0) * I_2(X_2, X_3|X_1 = 0)}$$

And if we make the calculations

$$I_3(X_1, X_2, X_3) = \exp(a_7) \quad \text{(A.SE.13)}$$

n-way interactions

By extending the definitions of checking for deviation from multiplicativity to the n-way interactions, we have that one should compare

$RR_{X_1+X_2+\cdots X_n+}$ vs $RR_{X_1+X_2-\cdots X_n-} * RR_{X_1-X_2+X_3-\cdots X_n-} * \ldots * RR_{X_1-X_2-\cdots X_{n-1}-X_n+}$

We can calculate $\mathrm{TotI}_n(X_1, X_2, \ldots, X_n)$ to check if there is any deviation from multiplicativity

$$\mathrm{TotI}_n(X_1, X_2, \ldots, X_n) = \frac{RR_{X_1+X_2+\cdots X_n+}}{RR_{X_1+X_2-\cdots X_n-} * RR_{X_1-X_2+X_3-\cdots X_n-} * \ldots * RR_{X_1-X_2-\cdots X_{n-1}-X_n+}} \quad \text{(A.SE.14)}$$

To check deviation from additivity, we have from equation (5) that we have to check

$$\mathrm{TotRERI}_n(X_1, X_2, \ldots, X_n) = RR_{X_1+X_2+\cdots X_n+} - RR_{X_1+X_2-\cdots X_n-} - RR_{X_1+X_2-\cdots X_n-} - \cdots - RR_{X_1-X_2-\cdots X_{n-1}-X_n+} + (n-1) \quad \text{(A.SE.15)}$$

We will generalize the inequalities we showed for 2-way interactions to the n-way interactions. Both of them hold when there is no qualitative interaction. They also hold if we relax the assumptions of qualitative interaction and we assume only that

$RR_{X_1+X_2-\cdots X_n-} > RR_{X_1-X_2-\cdots X_n-} (= 1)$ and

$RR_{X_1-X_2+X_3-\cdots X_n-} > RR_{X_1-X_2-\cdots X_n-} (= 1)$ and

…

$RR_{X_1-X_2-\cdots X_{n-1}-X_n+} > RR_{X_1-X_2-\cdots X_n-} (= 1)$

More specifically, we will prove

*1) When the effects of n risk factors are either multiplicative of super-multiplicative, then the effects will be super-additive.*

Proof: If $\mathrm{TotI}_n(X_1, X_2, \ldots, X_n) \geq 1$, we have that $\text{TotRERI}_n(X_1, X_2, \ldots, X_n) \geq RR_{X_1+X_2-\cdots X_n-} * \; RR_{X_1-X_2+X_3-\cdots X_n-} * \ldots * \; RR_{X_1-X_2-\cdots X_{n-1}-X_n+}$

$$-RR_{X_1+X_2-\cdots X_n-} - RR_{X_1-X_2+X_3-\cdots X_n-} - \cdots - RR_{X_1-X_2-\cdots X_{n-1}-X_n+} + (n-1)$$

Now we have to prove that

$RR_{X_1+X_2-\cdots X_n-} * \; RR_{X_1-X_2+X_3-\cdots X_n-} * \ldots * \; RR_{X_1-X_2-\cdots X_{n-1}-X_n+}$

$-RR_{X_1+X_2-\cdots X_n-} - RR_{X_1-X_2+X_3-\cdots X_n-} - \cdots - RR_{X_1-X_2-\cdots X_{n-1}-X_n+} + (n-1) > 0$ (A.SE.16)

We have that

$RR_{X_1+X_2-\cdots X_n-} * \; RR_{X_1-X_2+X_3-\cdots X_n-} * \ldots * \; RR_{X_1-X_2-\cdots X_{n-1}-X_n+}$

$-RR_{X_1+X_2-\cdots X_n-} - RR_{X_1-X_2+X_3-\cdots X_n-} - \cdots - RR_{X_1-X_2-\cdots X_{n-1}-X_n+} + (n-1) =$

1st step → remove $RR_{X_1+X_2-\cdots X_n-}$ from the equation

$= RR_{X_1+X_2-\cdots X_n-} * \; (RR_{X_1-X_2+X_3-\cdots X_n-} * \ldots * \; RR_{X_1-X_2-\cdots X_{n-1}-X_n+} - 1)$

$-RR_{X_1-X_2+\cdots X_n-} - \cdots - RR_{X_1-X_2-\cdots X_{n-1}-X_n+} + (n-1) >$ (because $RR_{X_1+X_2-\cdots X_n-} > 1$)

$(RR_{X_1-X_2+X_3-\cdots X_n-} * \ldots * \; RR_{X_1-X_2-\cdots X_{n-1}-X_n+} - 1)$

$-RR_{X_1-X_2+\cdots X_n-} - \cdots - RR_{X_1-X_2-\cdots X_{n-1}-X_n+} + (n-1) =$

$(RR_{X_1-X_2+X_3-\cdots X_n-} * \ldots * \; RR_{X_1-X_2-\cdots X_{n-1}-X_n+})$

$-RR_{X_1-X_2+\cdots X_n-} - \cdots - RR_{X_1-X_2-\cdots X_{n-1}-X_n+} + (n-2) =$

2nd step → remove $RR_{X_1-X_2+X_3-\cdots X_n-}$ from the equation

$= RR_{X_1-X_2+X_3-\cdots X_n-} * \; (RR_{X_1-X_2-X_3+X_4-\cdots X_n-} * \ldots * \; RR_{X_1-X_2-\cdots X_{n-1}-X_n+} - 1)$

$-RR_{X_1-X_2-X_3+X_4-\cdots X_n-} - \cdots - RR_{X_1-X_2-\cdots X_{n-1}-X_n+} + (n-2) >$ (because $RR_{X_1-X_2+X_3-\cdots X_n-} > 1$)

$(RR_{X_1-X_2-X_3+X_4-\cdots X_n-} * \ldots * \; RR_{X_1-X_2-\cdots X_{n-1}-X_n+} - 1)$

$-RR_{X_1-X_2+\cdots X_n-} - \cdots - RR_{X_1-X_2-\cdots X_{n-1}-X_n+} + (n-2) =$

$(RR_{X_1-X_2-X_3+X_4-\cdots X_n-} * \ldots * RR_{X_1-X_2-\cdots X_{n-1}-X_n+})$

$-RR_{X_1-X_2+\cdots X_n-} - \cdots - RR_{X_1-X_2-\cdots X_{n-1}-X_n+} + (n-3) >$

…

(n-2)th step → remove $RR_{X_1-X_2-\cdots X_{n-3}-X_{n-2}+X_{n-1}-X_n-}$ from the equation)

$RR_{X_1-X_2-\cdots X_{n-3}-X_{n-2}+X_{n-1}-X_n-} * (RR_{X_1-X_2-\cdots X_{n-2}-X_{n-1}+X_n-} * RR_{X_1-X_2-\cdots X_{n-1}-X_n+} - 1)$

$-RR_{X_1-X_2-\cdots X_{n-2}-X_{n-1}+X_n-} - RR_{X_1-X_2-\cdots X_{n-1}-X_n+} + (n-(n-2)) >$

(because $RR_{X_1-X_2-\cdots X_{n-3}-X_{n-2}+X_{n-1}-X_n-} > 1$)

$(RR_{X_1-X_2-\cdots X_{n-2}-X_{n-1}+X_n-} * RR_{X_1-X_2-\cdots X_{n-1}-X_n+} - 1)$

$-RR_{X_1-X_2-\cdots X_{n-2}-X_{n-1}+X_n-} - RR_{X_1-X_2-\cdots X_{n-1}-X_n+} + \left(n-(n-2)\right) =$

$(RR_{X_1-X_2-\cdots X_{n-2}-X_{n-1}+X_n-} * RR_{X_1-X_2-\cdots X_{n-1}-X_n+})$

$-RR_{X_1-X_2-\cdots X_{n-2}-X_{n-1}+X_n-} - RR_{X_1-X_2-\cdots X_{n-1}-X_n+} + 2 - 1 =$

$(RR_{X_1-X_2-\cdots X_{n-2}-X_{n-1}+X_n-} * RR_{X_1-X_2-\cdots X_{n-1}-X_n+})$

$-RR_{X_1-X_2-\cdots X_{n-2}-X_{n-1}+X_n-} - RR_{X_1-X_2-\cdots X_{n-1}-X_n+} + 1 =$

$\left(RR_{X_1-X_2-\cdots X_{n-2}-X_{n-1}+X_n-}\right) * \left(RR_{X_1-X_2-\cdots X_{n-1}-X_n+} - 1\right) - (RR_{X_1-X_2-\cdots X_{n-1}-X_n+} - 1) =$

$\left(RR_{X_1-X_2-\cdots X_{n-2}-X_{n-1}+X_n-} - 1\right) * \left(RR_{X_1-X_2-\cdots X_{n-1}-X_n+} - 1\right) > 0$

because $RR_{X_1-X_2-\cdots X_{n-2}-X_{n-1}+X_n-}$ *and* $RR_{X_1-X_2-\cdots X_{n-1}-X_n+} > 1$

So, we proved that when the effects of n risk factors are either multiplicative of super-multiplicative, then the effects will be super-additive.

*2) When the effects of n risk factors are either additive or sub-additive, then the effects will be sub-multiplicative.*

Proof: If $\mathrm{TotRERI_n}(X_1, X_2, \dots, X_n) \leq 0$, then

$$RR_{X_1+X_2-\cdots X_n-} * \; RR_{X_1-X_2+X_3-\cdots X_n-} \; * \dots * \; RR_{X_1-X_2-\cdots X_{n-1}-X_n+}$$

$$-RR_{X_1+X_2-\cdots X_n-} - RR_{X_1-X_2+X_3-\cdots X_n-} - \cdots - RR_{X_1-X_2-\cdots X_{n-1}-X_n+} + (n-1) \leq 0$$

or equivalently

$$\frac{RR_{X_1+X_2-\cdots X_n-} * \; RR_{X_1-X_2+X_3-\cdots X_n-} * \dots * \; RR_{X_1-X_2-\cdots X_{n-1}-X_n+}}{RR_{X_1+X_2-\cdots X_n-} + RR_{X_1-X_2+X_3-\cdots X_n-} + \cdots + RR_{X_1-X_2-\cdots X_{n-1}-X_n+} - (n-1)} \leq 1 \quad \text{(A.SE.17)}$$

From (A.SE.14), we have that

$$\mathrm{TotI}_n(X_1, X_2, \dots, X_n) = \frac{RR_{X_1+X_2+\cdots X_n+}}{RR_{X_1+X_2-\cdots X_n-} * \; RR_{X_1-X_2+X_3-\cdots X_n-} \; * \dots * \; RR_{X_1-X_2-\cdots X_{n-1}-X_n+}} <$$

(because of (A.SE.16))

$$\frac{RR_{X_1+X_2-\cdots X_n-} * \; RR_{X_1-X_2+X_3-\cdots X_n-} * \dots * \; RR_{X_1-X_2-\cdots X_{n-1}-X_n+}}{RR_{X_1+X_2-\cdots X_n-} + RR_{X_1+X_2-\cdots X_n-} + \cdots + RR_{X_1-X_2-\cdots X_{n-1}-X_n+} - (n-1)} \leq 1 \quad \text{because of (A.SE.17)}$$

So we proved that when $\mathrm{TotRERI_n}(X_1, X_2, \dots, X_n) \leq 0$, then $\mathrm{TotI}_n(X_1, X_2, \dots, X_n) < 1$. In other words, we showed that when the effects of n risk factors are either additive or sub-additive, then the effects will be sub-multiplicative.

<u>Connection between deviation from additivity and multiplicativity from the worked example</u>

We can calculate $\mathrm{TotI_3}$=1.20>1 from (A.SE.6), so the effects of low MD, obesity and smoking status on mortality are super-multiplicative (even not statistically significant), meaning that we also expect that these effects would also be super-additive (from the 1st inequality). This is true, because $\mathrm{TotRERI_3}$=1.20>0, in other words, there is an excess 120% risk due to the joint presence of all risk factors, compared to the situation that each of them would act separately

The 3-way interaction of these factors beyond the 2-way interactions was positive both under the additive and under the multiplicative scale ($RERI_3$=1.98 and $I_3$=2.51), however there is not a direct link between these two indeces. The only conclusion that we can infer from equations (A.SB3) and (A.SE13) is that the greater the value of $I_3$ is, the greater the value of $RERI_3$ will be as well (the opposite is not always true), without any guarantee that $RERI_3$ will be positive, depending on a specific value for $I_3$.

Moreover, all the $RERI_2$ given the absence of the 3$^{rd}$ risk factor are negative, indicating that the corresponding all the effects between 2 risk factors, when the 3$^{rd}$ is absent, will be sub-multiplicative (from the 2$^{nd}$ inequality). More specifically, we had that

$I_2$(low MD, high BMI / never or former smokers)=0.77<1

$I_2$(low MD, current smokers / low BMI)= 0.79<1

$I_2$(high BMI, current smokers / high MD)= 0.79<1

Regarding the 2-way interactions, when the 3$^{rd}$ factor was present, when we calculate that all the $I_2$'s, we find that the effects of every 2 risk factors, when the 3$^{rd}$ is present, were super-multiplicative;

$I_2$(low MD, high BMI / current smokers)=1.98>1

$I_2$(low MD, current smokers / high BMI)= 1.99>1

$I_2$(high BMI, current smokers / low MD)= 1.92>1

indicating that the corresponding effects will be super-additive, that's why all $RERI_2$'s were positive.

In this example, we found that the interpretation doesn't change, if we convert to deviation from multiplicativity as reference. However, this is not always true. We showed above that

1) If $TotI_3 \geq 1 \rightarrow TotRERI_3 > 0$ and

2) If $TotRERI_3 \leq 0 \rightarrow TotI_3 < 1$

The opposite in these relationships does not always hold. In other words, it is possible to observe super-additive effects, which can be sub-multiplicative ($TotRERI_3>0$ & $TotI_3<1$)

We also mentioned that the greater the value of $I_3$ is, the greater the value of $RERI_3$ will be. Nevertheless, there is no specific interval lower limit for $RERI_3$ for different values of $I_3$.

**Section F – Qualitative interaction**

We refer to the term qualitative (or cross-over) interaction when the exposure of interest is a risk factor for a specific outcome for one subgroup, but a protective factor for another subgroup. For example, a specific medication might be beneficial in younger people, but not in the elderly. Qualitative interaction is very crucial for decision making for public health purposes, because, in such instances, we should not treat all the subgroups, but only those people for which the medication is beneficial.

To check for qualitative interaction in case of 2 risk factors $X_1$ and $X_2$, we should check

1) Whether the risk of $X_1$ is increasing across the strata of $X_2$, i.e.

$RR_{X_1+X_2+} > RR_{X_1-X_2+}$ and $RR_{X_1+X_2-} > RR_{X_1-X_2-}$

and

2) Whether the risk of $X_2$ is increasing across the strata of $X_1$, i.e.

$RR_{X_1+X_2+} > RR_{X_1+X_2-}$ and $RR_{X_1-X_2+} > RR_{X_1-X_2-}$

To apply the same in case of 3 risk factors $X_1$, $X_2$ and $X_3$, we should check

1) whether the risk of $X_1$ is increasing across the strata of $X_2$ and $X_3$, i.e.

$RR_{X_1+X_2+X_3+} > RR_{X_1-X_2+X_3+}$ , $RR_{X_1+X_2+X_3-} > RR_{X_1-X_2+X_3-}$ , $RR_{X_1+X_2-X_3+} > RR_{X_1-X_2-X_3+}$

and $RR_{X_1+X_2-X_3-} > RR_{X_1-X_2-X_3-}$

2) whether the risk of $X_2$ is increasing across the strata of $X_1$ and $X_3$, i.e.

$RR_{X_1+X_2+X_3+} > RR_{X_1+X_2-X_3+}$ , $RR_{X_1+X_2+X_3-} > RR_{X_1+X_2-X_3-}$ , $RR_{X_1-X_2+X_3+} > RR_{X_1-X_2-X_3+}$

and $RR_{X_1-X_2+X_3-} > RR_{X_1-X_2-X_3-}$

3) whether the risk of $X_3$ is increasing across strata of $X_1$ and $X_2$, i.e.

$RR_{X_1+X_2+X_3+} > RR_{X_1+X_2+X_3-}$ , $RR_{X_1+X_2-X_3+} > RR_{X_1+X_2-X_3-}$ , $RR_{X_1-X_2+X_3+} > RR_{X_1-X_2+X_3-}$ and $RR_{X_1-X_2-X_3+} > RR_{X_1-X_2-X_3-}$

The same procedure should be followed for multi-way interactions, i.e. to check

1) whether the risk of $X_1$ is increasing across the strata of $X_2$, $X_3$ … $X_n$,.

2) whether the risk of $X_2$ is increasing across the strata of $X_1$, $X_3$ … $X_n$,

…

n) whether the risk of $X_n$ is increasing across the strata of $X_1$, $X_2$ … $X_{n-1}$

In Section C of the Appendix, we show how these calculations can be applied in Stata, in case of 3 risk factors, $X_1$, $X_2$ and $X_3$.

Qualitative interaction occurs very rarely in the study of the joint effects of 2 risk factors, that's why it is not mentioned in the literature frequently. It is very likely to happen rarely in the case of 3 risk factors as well, but it is more possible to occur when studying more >3 risk factors (the more we increase the factors of interest, the more likely to observe qualitative interactions)
So, when conducting a multi-way interaction analysis, we should additionally test whether the risk of a risk factor, is increasing across the different subgroups of interest of the population.
In other words, a researcher should assess the results from the interaction analysis (TotRERI and RERIs), as we did in the main body of the manuscript, and additionally test whether there is qualitative interaction. In the case that qualitative interaction exists, then one should comment on the corresponding consequences (e.g. that a specific medication is protective for CVD in one subgroup of the study while it is not in another), apart from the discussion of the results of interaction analysis (RERIs and TotRERIs).